\documentclass[aps,prb,reprint,superscriptaddress]{revtex4-1}
\usepackage{amsmath}
\usepackage{bm}
\usepackage{multirow}
\usepackage{graphicx}
\usepackage{epstopdf}
\usepackage{tabularx, booktabs}
\usepackage{pbox}
\newcolumntype{Y}{>{\centering\arraybackslash}X}
\usepackage[usenames,dvipsnames]{color}
\usepackage[caption=false,position=b,singlelinecheck=off,font=normalsize,labelfont=bf,justification=justified]{subfig}
\usepackage{hyperref}

\hypersetup{
    colorlinks=true,
    linkcolor=blue,
    filecolor=black,      
    urlcolor=blue,
    citecolor=blue,
}
\graphicspath{ {Figures_PDF/} }

\begin{document}

\title{Investigation into the role of the orbital moment in a series of isostructural weak ferromagnets}

\author{D. Pincini}
\email[]{davide.pincini.14@ucl.ac.uk}
\affiliation{London Centre for Nanotechnology and Department of Physics and Astronomy, University College London, Gower Street, London WC1E6BT, UK}
\affiliation{Diamond Light Source Ltd., Diamond House, Harwell Science \& Innovation Campus, Didcot, Oxfordshire OX11 0DE, UK}

\author{F. Fabrizi}
\affiliation{Diamond Light Source Ltd., Diamond House, Harwell Science \& Innovation Campus, Didcot, Oxfordshire OX11 0DE, UK}

\author{G. Beutier}
\affiliation{Univ. Grenoble Alpes, CNRS, Grenoble INP, SIMaP, F-38000 Grenoble, France}

\author{G. Nisbet}
\affiliation{Diamond Light Source Ltd., Diamond House, Harwell Science \& Innovation Campus, Didcot, Oxfordshire OX11 0DE, UK}

\author{H. Elnaggar}
\affiliation{Utrecht University, Debye Institute for Nanomaterials Science, Universiteitsweg 99, 3584 CG, Utrecht, The Netherlands }

\author{V.E. Dmitrienko}
\affiliation{A.V. Shubnikov Institute of Crystallography, FSRC "Crystallography and Photonics" RAS, Moscow 119333, Russia}

\author{M.I. Katsnelson}
\affiliation{Radboud University Nijmegen, Institute for Molecules and Materials, Heyendaalseweg 135, NL-6525 AJ Nijmegen, The Netherlands}
\affiliation{Department of Theoretical Physics and Applied Mathematics, Ural Federal University, Mira str. 19, 620002 Ekaterinburg, Russia}

\author{Y.O. Kvashnin}
\affiliation{Department of Physics and Astronomy, Division of Materials Theory, Uppsala University, Box 516, SE-75120 Uppsala, Sweden}

\author{A.I. Lichtenstein}
\affiliation{I. Institut f\"{u}r Theoretische Physik, Universität Hamburg, Jungiusstraÿe 9, D-20355 Hamburg, Germany}
\affiliation{Department of Theoretical Physics and Applied Mathematics, Ural Federal University, Mira str. 19, 620002 Ekaterinburg, Russia}

\author{V.V. Mazurenko}
\affiliation{Department of Theoretical Physics and Applied Mathematics, Ural Federal University, Mira str. 19, 620002 Ekaterinburg, Russia}

\author{E.N. Ovchinnikova}
\affiliation{Faculty of Physics, M.V.Lomonosov Moscow State University, Leninskie Gory, Moscow 119991, Russia}

\author{O.V. Dimitrova}
\affiliation{Faculty of Physics, M.V.Lomonosov Moscow State University, Leninskie Gory, Moscow 119991, Russia}

\author{S.P. Collins}
\affiliation{Diamond Light Source Ltd., Diamond House, Harwell Science \& Innovation Campus, Didcot, Oxfordshire OX11 0DE, UK}

\date{\today}

\begin{abstract}
The orbital contribution to the magnetic moment of the transition metal ion in the isostructural weak ferromagnets ACO$_3$ (A=Mn,Co,Ni) and FeBO$_3$ was investigated by a combination of first-principles calculations, non-resonant x-ray magnetic scattering and x-ray magnetic circular dichroism. A non-trivial evolution of the orbital moment as a function of the $3d$ orbitals filling is revealed, with a particularly large value found in the Co member of the family. Here, the coupling between magnetic and lattice degrees of freedom produced by the spin-orbit interaction results in a large single-ion anisotropy and a peculiar magnetic-moment-induced electron cloud distortion, evidenced by the appearance of a subtle scattering amplitude at space group-forbidden reflections and significant magnetostrictive effects. Our results, which complement a previous investigation on the sign of the Dzyaloshinskii$-$Moriya interaction across the series, highlight the importance of spin-orbit coupling in the physics of weak ferromagnets and prove the ability of modern first-principles calculations to predict the properties of materials where the Dzyaloshinskii$-$Moriya interaction is a fundamental ingredient of the magnetic Hamiltonian.
\end{abstract}

\maketitle

\section{Introduction}

Recent studies of the weak ferromagnetic carbonates ACO$_3$ (A=Mn,Co,Ni)\cite{Beutier2017} and FeBO$_3$ \cite{dmitrienko_measuring_2014} represent the first systematic experimental and theoretical investigation of the changes in the sign and magnitude of the Dzyaloshinskii$-$Moriya interaction (DMI) across a series of insulating $3d$ transition metal (TM) compounds. The combination of novel resonant x-ray diffraction technique and modern first-principles calculations revealed a dramatic evolution of the sign of the DMI as the $3d$ orbitals of the TM are gradually filled with electrons. The ability to accurately model the DMI is essential for the fundamental understanding of a plethora of exotic non-collinear magnetic ground states, such as spin spirals\cite{Bode2007} and Skyrmions \cite{yu_real-space_2010,Heinze2011,Huang2012}, and their exploitation as candidate materials for spintronics applications.

The DMI has its microscopic origin in spin-orbit coupling (SOC) \cite{moriya_anisotropic_1960,Moriya1960}. In the common paradigm of the physics of TM oxides, SOC is regarded as negligible for $3d$ electrons, where its role is merely as a small perturbation to the ground-state Hamiltonian \cite{khomskii_transition_2014}. This contrasts with the case of heavier ($4d$ and $5d$) TM compounds, where SOC competes with the crystal field and other relevant energy scales on an equal footing and gives rise to more exotic ground states \cite{witczak-krempa_correlated_2014}. Nonetheless, even for $3d$ TM compounds, SOC is expected to have a significant impact on the magnetic properties of the system whenever a finite orbital moment is present\cite{khomskii_transition_2014}. A substantial unquenched orbital contribution to the magnetic moment has been indeed reported for several $3d$ oxides \cite{Fernandez1998a,Neubeck1999,Kwon2000,Neubeck2001,Lee2016,Huang2004,Okamoto2014,Radwanski2004}. In this case, the coupling between spin and orbital moment caused by the spin-orbit interaction can generally produce a strong magnetoelastic coupling and lead to the appearance of a large single-ion anisotropy and magnetostrictive effects\cite{khomskii_transition_2014}. The magnetic properties will then considerably differ from the case a spin-only system with quenched orbital degrees of freedom.


A careful determination of the strength of the orbital moment and its impact on the magnetic ground state is of particular interest for the weak ferromagnets ACO$_3$ (A=Mn,Co,Ni) and FeBO$_3$, where SOC, and the resulting DMI, underpins one of the most peculiar aspects of the physics of this system, i.e. the existence of a weak net magnetization. 
Weak ferromagnets also represent an ideal model system for (i) their manageable magnetic unit cell, in contrast to the more complex spin-spiral and Skyrmion states of interest in light of spintronics applications and (ii) the availability of state of the art calculations \cite{Beutier2017,dmitrienko_measuring_2014} which can be conveniently used to predict the relative orbital and spin contribution to the TM magnetic moment. 

In this paper we present a detailed investigation into the role of the orbital moment in the isostructural weak ferromagnets ACO$_3$ (A=Mn,Co,Ni) and FeBO$_3$ by means of a combination of theoretical calculations, Non-resonant X-ray Magnetic Scattering (NXMS) and X-ray Magnetic Circular Dichroism (XMCD). While MnCO$_3$ and FeBO$_3$ behave as almost pure spin systems, a sizeable orbital contribution to the magnetic moment was found in CoCO$_3$ and NiCO$_3$. In particular, a large orbital moment is present in the Co compound which results in a remarkable coupling between lattice and magnetic degrees of freedom. The latter is unveiled by a sizeable magnetocrystalline anisotropy of the magnetic interactions and, more spectacularly, by the emergence of an unusual space-group-forbidden scattering process. 

The paper is organized as follows. A brief description of the samples is given in \textbf{\S}~\ref{sec:samples}, while the theoretical calculations and the NXMS and XMCD experimental setup are outlined in \textbf{\S}~\ref{sec:methods}. \textbf{\S}~\ref{sec:systematic_investigation} presents the results of the DFT calculations and the NXMS measurements on the orbital contribution to the magnetic moment across the series. \textbf{\S}~\ref{sec:CoCO3} includes specific findings on CoCO$_3$, in particular: \textbf{\S}~\ref{sec:CoCO3_XMCD} outlines the XMCD measurements used to support the NXMS results on the size of the orbital moment, \textbf{\S}~\ref{sec:anisotropy} discusses the role of the magnetocrystalline anisotropy in the NXMS data and \textbf{\S}~\ref{sec:forbidden_scattering} deals with the space-group forbidden scattering and its microscopical interpretation based on the multiplet calculations. Finally, the concluding remarks are presented in \textbf{\S}~\ref{sec:conclusions}.

\section{Samples}\label{sec:samples}

\begin{figure}
	\centering
	\includegraphics[width=0.9\linewidth]{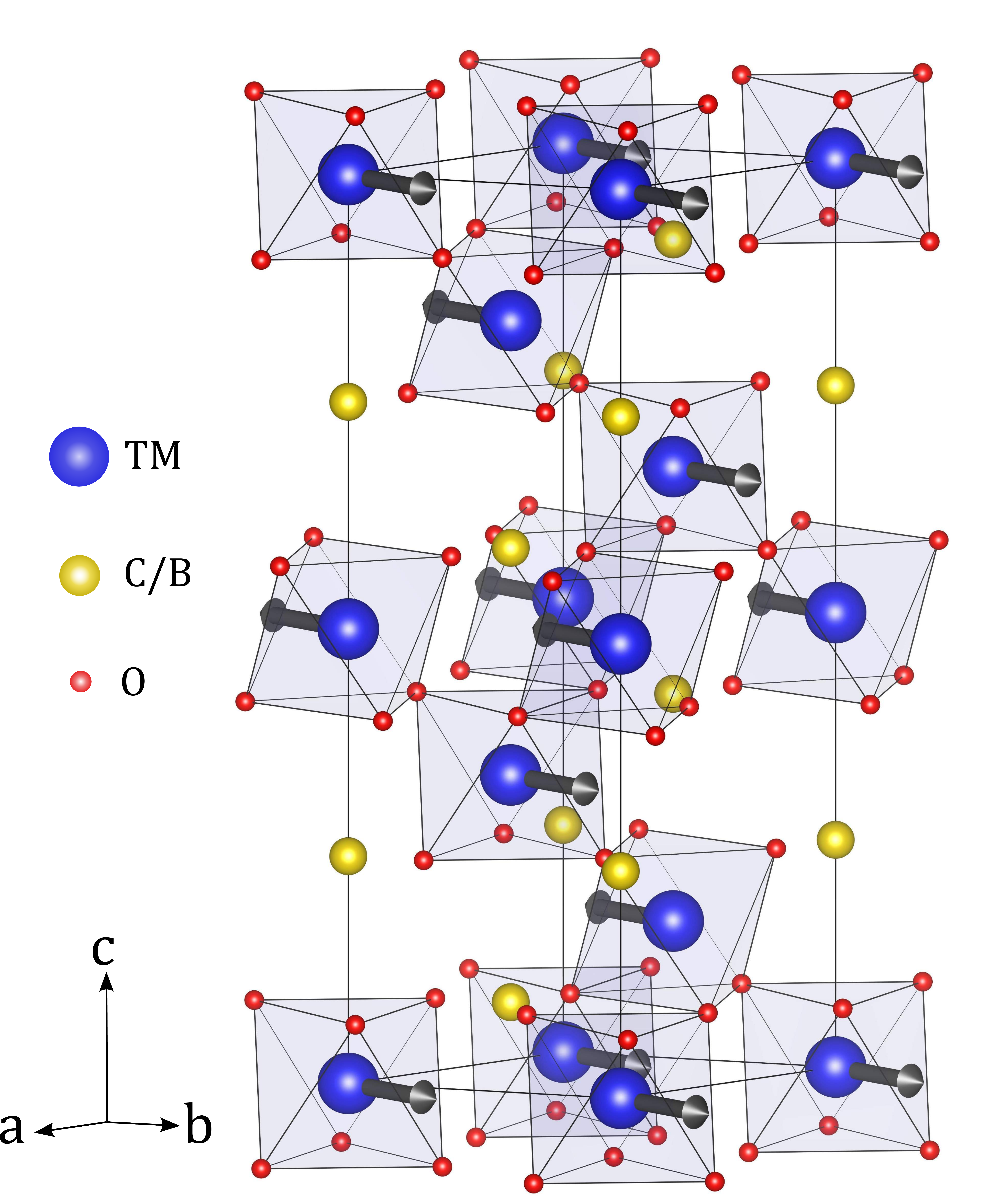}
	\caption{(Color online) Trigonal crystal structure (space group R$\bar{3}$c, No.~167) of the weak ferromagnets ACO$_3$ (A=Mn,Co,Ni) and FeBO$_3$ in the hexagonal axes description. Large blue spheres: transition metal (TM) atoms (Mn, Fe, Co, Ni); medium-size yellow spheres: C/B; small red spheres: O. The arrows represent the magnetic moments of the TM atoms in the AFM phase.}
	\label{CoCO3_crystal_structure}
\end{figure}

The weak ferromagnetic carbonates ACO$_3$ (A=Mn,Co,Ni) and FeBO$_3$ are isostructural compounds, with the trigonal R$\bar{3}$c crystal symmetry \cite{Pertlik1986,Maslen1995,Diehl1975a,Wyckoff1920} (Fig.~\ref{CoCO3_crystal_structure}). The latter consists of alternating TM and oxygen-carbon/boron layers, such that each TM ion is at the center of a distorted TMO$_6$ octahedra. The TM magnetic moments of ACO$_3$ (FeBO$_3$) display an analogous antiferromagnetic (AFM) order at low (room) temperature: the moments lie in the crystal $\mathbf{ab}$ basal plane, and are coupled ferromagnetically in each TM layer and antiferromagnetically between adjacent layers. The moments in different layers, however, are not exactly antiparallel one to another: the finite DMI causes the moments to be slightly canted and results in a small net magnetization in the basal plane of the crystal \cite{Beutier2017,Dmitrienko2010,dmitrienko_measuring_2014} (Fig.~\ref{CoCO3_crystal_structure}).
The single crystals used in the present investigation are the same as Ref.~\onlinecite{Beutier2017,dmitrienko_measuring_2014}, which the reader is referred to for further details on the crystal and magnetic structures and the sample growth.

\section{Methods}\label{sec:methods}

\bgroup
\footnotesize
\def\arraystretch{1.2}
\begin{table*}[htp]
	\centering
	
	\resizebox{\textwidth}{!}{
		\begin{tabular}{c|c|c|c|c|ccccccc|c}
			\hline\hline
			\multirow{2}{*}{Compound} & \multirow{2}{*}{\begin{tabular}[c]{@{}c@{}}Magnetic \\ ion\end{tabular}} & \multirow{2}{*}{   $Z$   } & \multirow{2}{*}{\begin{tabular}[c]{@{}c@{}}$3d$ \\ electrons\end{tabular}} & \multirow{2}{*}{\begin{tabular}[c]{@{}c@{}}AFM\\ sublattice\end{tabular}} & \multicolumn{7}{c|}{Calculated spin and orbital angular momenta}                                  & \multirow{2}{*}{\begin{tabular}[c]{@{}c@{}}Measured\\ $\lvert \mathbf{l}\rvert/\lvert \mathbf{s}\rvert$\end{tabular}} \\ \cline{6-12}
			&                                                                          &                      &                                                                            &                                                                           & $s_x$  & $s_y$  & $s_z$  & $l_x$  & $l_y$  & $l_z$  & $\lvert \mathbf{l}\rvert/\lvert \mathbf{s}\rvert$ &                                                                                                                 \\ \hline\hline
			\multirow{2}{*}{MnCO$_3$} & \multirow{2}{*}{Mn$^{2+}$}                                               & \multirow{2}{*}{25}  & \multirow{2}{*}{5.0}                                                       & A                                                                         & 2.252  & 0.002  & 0      & 0.002      & 0      & 0      & \multirow{2}{*}{0.0009}                          & \multirow{2}{*}{0.05(2)}                                                                                        \\
			&                                                                          &                      &                                                                            & B                                                                         & -2.252 & 0.002  & 0      & -0.002      & 0      & 0      &                                             &                                                                                                                 \\ \hline
			\multirow{2}{*}{FeBO$_3$} & \multirow{2}{*}{Fe$^{3+}$}                                               & \multirow{2}{*}{26}  & \multirow{2}{*}{5.8}                                                       & A                                                                         & 2.059  & 0.029  & 0      & 0.021  & 0      & 0      & \multirow{2}{*}{0.010}                      & \multirow{2}{*}{0.03(2)}                                                                                        \\
			&                                                                          &                      &                                                                            & B                                                                         & -2.059 & 0.029  & 0      & -0.021 & 0      & 0      &                                             &                                                                                                                 \\ \hline
			\multirow{2}{*}{CoCO$_3$} & \multirow{2}{*}{Co$^{2+}$}                                               & \multirow{2}{*}{27}  & \multirow{2}{*}{7.1}                                                       & A                                                                         & 1.289  & -0.108 & -0.010 & 0.736  & -0.058 & -0.004 & \multirow{2}{*}{0.57}                       & \multirow{2}{*}{0.7(2)}                                                                                         \\
			&                                                                          &                      &                                                                            & B                                                                         & -1.289 & -0.108 & 0.010  & -0.736 & -0.058 & 0.004  &                                             &                                                                                                                 \\ \hline
			\multirow{2}{*}{NiCO$_3$} & \multirow{2}{*}{Ni$^{2+}$}                                               & \multirow{2}{*}{28}  & \multirow{2}{*}{8.2}                                                       & A                                                                         & 0.801  & -0.105 & 0      & 0.190  & -0.024 & 0      & \multirow{2}{*}{0.24}                       & \multirow{2}{*}{0.3(2)}                                                                                         \\
			&                                                                          &                      &                                                                            & B                                                                         & -0.801 & -0.105 & 0      & -0.190 & -0.024 & 0      &                                             &                                                                                                                 \\ \hline\hline
		\end{tabular}
	}
	\caption{Spin and orbital angular momenta in units of $\hbar$ for the different compounds of the series A(C,B)O$_3$ (A=Mn,Fe,Co,Ni) as derived from DFT calculations and measured by means of NXMS. The $\mathbf{xyz}$ reference frame is defined such that $\mathbf{x}$ is perpendicular to a 2-fold axis and contained in a $\mathbf{c}$ glide plane of the R$\bar{3}$c structure and $\mathbf{z}$ is parallel to the crystallographic $\mathbf{c}$ axis.}
	\label{tab:calculations}
\end{table*}
\egroup

\subsection{First-principles calculations}\label{sec:calculations}
The orbital and spin moments of the selected compounds were calculated using the Vienna ab initio simulation package (VASP) \cite{Kresse1996,Kresse1999} within the local density approximation taking into account the on-site Coulomb interaction $U$ and SOC (LDA $+$ $U$ $+$ SO) \cite{Solovyev1998}. The calculations are the same as outlined in our recent resonant scattering investigation \cite{Beutier2017}, where further details on the calculation methods can be found. The initial magnetisation directions were set to lie along the $\mathbf{x}$ direction, with $\mathbf{x}$ perpendicular to a 2-fold axis and contained in a $\mathbf{c}$ glide plane of the R$\bar{3}$c structure. This results in having a canted AFM state, which is the lowest-energy state for all compounds. The results will be discussed and compared to the experiment in \textbf{\S}~\ref{sec:systematic_investigation}. Values of spin and
orbital moments reported in the present work are projections of the magnetisation density onto a sphere around the corresponding TM ion. Due to covalent bonding of the TM $3d$ orbitals with the oxygens $2p$ states, part of the magnetisation density appears on the ligand sites. The latter also contributes to the net magnetic moment.

\begin{figure*}[htp]
	\centering
	\includegraphics[width=\textwidth]{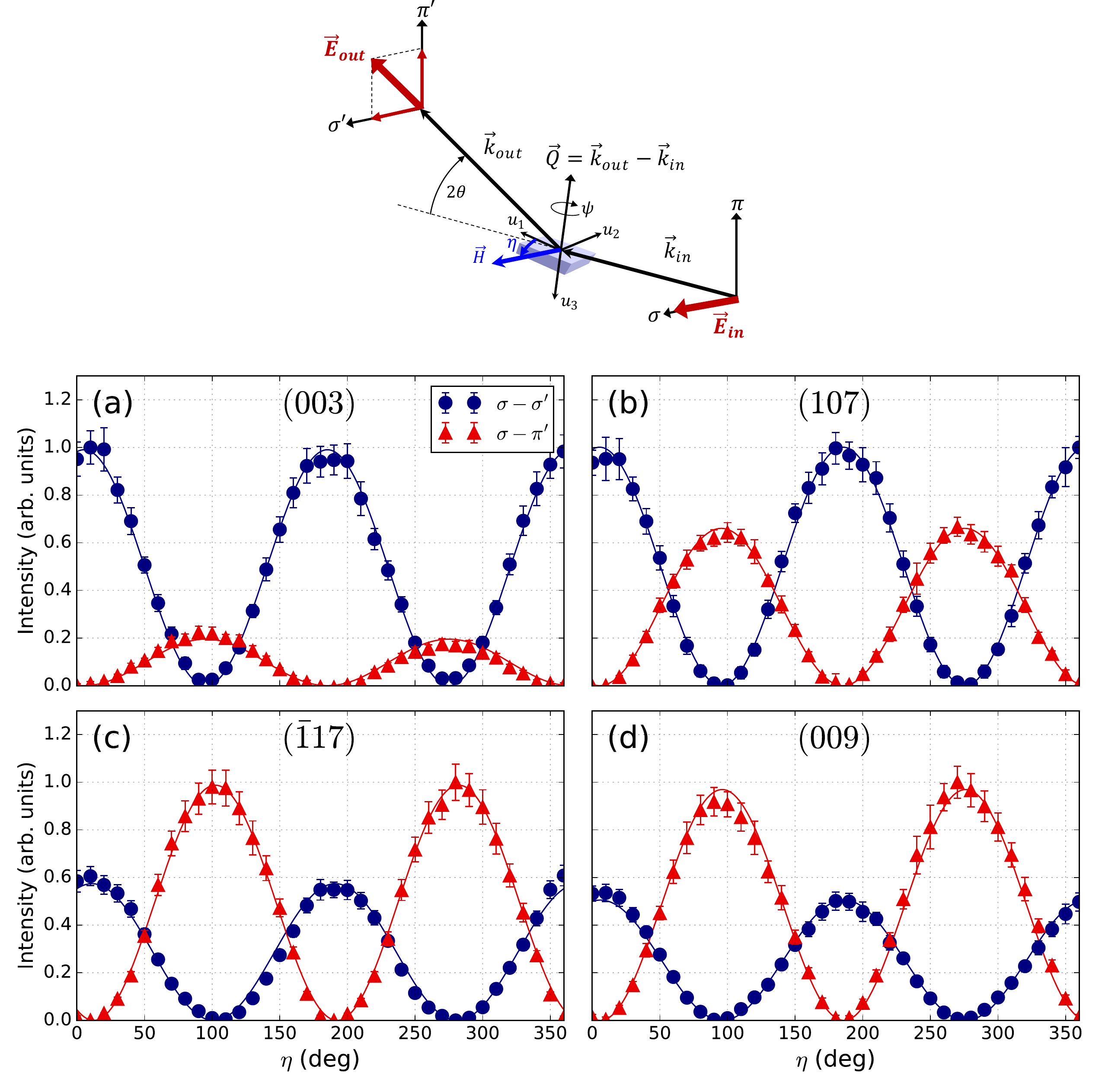}
	\caption[]{(Color online) Representative magnetic reflections dependence on the magnetic field direction for two different polarization states of the diffracted x-ray beam in CoCO$_3$. The data points represent the diffracted intensity integrated over a rocking scan while the solid curves correspond to the best fit to Eq.~(\ref{eq:intensities}). The data were collected at $T=5-6$~K and $\psi=83^\circ,0^\circ,108^\circ,30^\circ$ for the $(003)$, $(107)$, $(\bar{1}17)$ and $(009)$ reflection, respectively. A small constant background originating from residual multiple scattering has been removed from all the data sets. For each reflection, the intensity is normalized to the maximum value across both polarization channels. A schematic drawing of the vertical scattering geometry, along with the definition of the $\mathbf{u}_1\mathbf{u}_2\mathbf{u}_3$ reference frame used to express the cross sections of Eq.~(\ref{eq:scattering_amplitudes}) and the magnetic field angle $\eta$, is reported in the top panel.}
	\label{CoCO3_oscillations}
\end{figure*}

\subsection{Non-resonant X-ray Magnetic Scattering}\label{sec:experimental_NXMS}
The NXMS \cite{DeBergevin1981,blume_polarization_1988} measurements were performed in vertical scattering geometry at beamline I16 of the Diamond Light Source, Didcot UK \cite{Collins2010a}. The crystals were mounted on the sample rotational stages of the 6-circle kappa diffractometer with the $\mathbf{c}$ axis of the R$\bar{3}$c trigonal structure aligned vertically. A standard closed-cycle cryostat was used to cool down the ACO$_3$ samples below the N\'{e}el transition temperature of the canted AFM structure, while the data on FeBO$_3$ were collected at room temperature. A magnetic field $\mu_0 H\approx35$~mT, sufficient to drive the canted AFM structure into a single-domain phase \cite{Borovik-Romanov1961a,supplemental_material}, was applied to the $\mathbf{ab}$ plane of the crystal using the rotating permanent magnet setup already successfully employed in our recent resonant x-ray scattering measurements \cite{dmitrienko_measuring_2014,Beutier2017}. 

The diffracted signal arising from several space-group forbidden reflections of the type $(\text{00L}), \text{L}=6n+3$ and $(\text{H}\bar{\text{H}}\text{L}), \text{L}=2n+1$ was measured using linearly polarized radiation ($30 \times 200\,\mu m^2$ spot size), with the electric field vector of the incident x-rays lying in the horizontal plane (referred to as $\sigma$ polarization following the conventions by \citet{blume_polarization_1988}). For each reflection, the scattered signal for both the rotated ($\sigma-\pi'$) and unrotated ($\sigma-\sigma'$) polarization channels was measured as a function of the magnetic field direction (where $\sigma'$ and $\pi'$ denote the polarization of a scattered beam whose electric field vector is perpendicular or parallel to the scattering plane, respectively). The latter is described by the angle $\eta$: this is defined such that the field lies in the vertical scattering plane (pointing towards the detector) for $\eta=0^\circ$ and is perpendicular to the latter for $\eta=90,270^\circ$. 

Polarization analysis of the scattered beam was achieved by means of the $(004)$ reflection from a pyrolytic graphite (PG) single crystal [with the exception of the data presented in \textbf{\S}~\ref{sec:anisotropy}, for which the $(006)$ was used]. The total scattered intensity without polarization analysis was also measured for MnCO$_3$, FeBO$_3$ and CoCO$_3$ in order to correct for the different reflection efficiencies of the PG crystal in the $\sigma-\pi'$ and $\sigma-\sigma'$ polarization channels (see Appendix~\ref{sec:data_treatment}). The energy of the incident beam was kept fixed to $E=5.223$~keV ($E=7.684$~keV for the data of \textbf{\S}~\ref{sec:anisotropy}), chosen for being away from any sample absorption edges and for minimizing the cross-talk between the two orthogonal light polarizations. For most of the measured reflections, equivalent data sets were collected at several different sample azimuths $\psi$\cite{supplemental_material}, whose values were selected to minimize the contribution of multiple scattering to the diffracted intensity. The corresponding results were then averaged. All the $\psi$ values reported in this paper are defined with respect to the $(100)$ azimuthal reference [$\psi=0^\circ$ when the $(100)$ reciprocal direction lies in the scattering plane pointing towards the detector].

\subsection{X-ray Magnetic Circular Dichroism}\label{sec:experimental_XMCD}
XMCD was measured on a single crystal of CoCO$_3$ at the high-field magnet end station (BLADE) of beamline I10 of the Diamond Light Source. A thin film of Pt ($\approx$ 2 nm) was deposited via sputter coating on the crystal's facet orthogonal to the $\mathbf{c}$ axis at the Research Complex at Harwell (Didcot, UK) prior to the XMCD measurements. The purpose of the Pt coating was to create an electrical contact with the illuminated area of the sample. The latter allowed the drain current of photo electrons to be extracted thus making total electron yield (TEY) detection possible despite the strong insulating character of CoCO$_3$. The crystal was clamped on a electrically-grounded copper holder and inserted in the UHV sample environment of the I10 superconducting magnet with the coated surface facing the incident beam. 

The measurements were performed at a shallow ($20^{\circ}$) incident x-ray angle, so that the external magnetic field, directed along the incident beam wave vector, was almost perpendicular to the $\mathbf{c}$ axis. A relatively small field value ($\mu_0 H=0.4$~T) was used: this was chosen to be sufficiently large to suppress the magnetic domains structure and thus generate a net magnetization along the field while being, at the same time, small enough not to significantly perturb the in-plane canted AFM order of the Co$^{2+}$ moments. X-ray Absorption Spectroscopy (XAS) measurements were collected across the Co $L_3$ (778.2 eV) and $L_2$ (793.1 eV) edges for opposite helicities of the incident circularly-polarized soft x-ray beam ($20 \times 100\,\mu m^2$ spot size) and opposite directions of the external field. Several XAS spectra were collected and averaged for each permutation of light polarization and field direction and the resulting spectra combined to obtain the XMCD signal. 

\section{Evolution of the orbital moment in (C\lowercase{o},N\lowercase{i},M\lowercase{n})CO$_3$ and F\lowercase{e}BO$_3$ }\label{sec:systematic_investigation}

\begin{figure}[htp]
	\centering
	\includegraphics[width=0.49\textwidth]{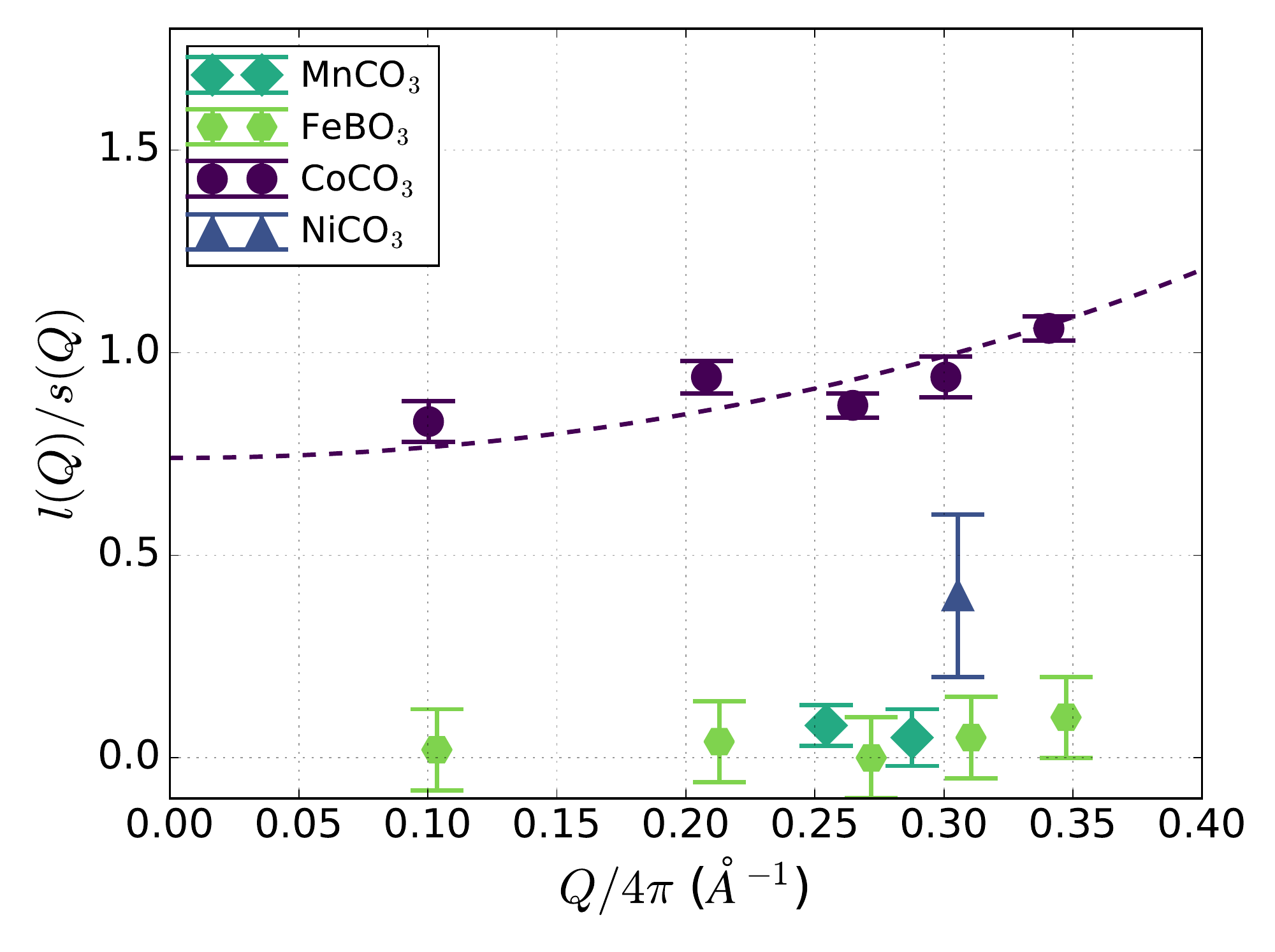}
	\caption[]{(Color online) Orbital-to-spin angular momenta ratio for the compounds of the series A(C,B)O$_3$ (A=Mn,Fe,Co,Ni) as a function of the momentum transfer. For each compound, the $l(Q)/s(Q)$ value for each magnetic reflection (and the corresponding errorbar) was calculated combining the measurements at different temperatures and azimuth values \cite{supplemental_material}. The dashed line represents a fit to Eq.~(\ref{eq:ratio}) of the CoCO$_3$ data considering an isotropic approximation of the magnetic form factors.}
	\label{Orbital_moment}
\end{figure}

The orbital and spin angular momenta derived from the DFT calculations are summarized in Table~\ref{tab:calculations}. The calculations predict a negligible orbital contribution to the total angular momentum of the Mn$^{2+}$ and Fe$^{3+}$ ions. On the other hand, a significant orbital angular momentum is found in CoCO$_3$ and NiCO$_3$. The orbital angular momentum is particularly large for the Co$^{2+}$ ion, where it reaches almost $60\%$ of the spin value. This peculiar trend does not find a trivial explanation in a simple isolated-ion picture. Although Hund's coupling applied to Mn$^{2+}$ and Fe$^{3+}$ ($3d^5$, $l=0,\,s=5/2$) predicts a zero orbital moment, the orbital contribution should be larger in Ni$^{2+}$ ($3d^8$, $l=3,\,s=1$) than in Co$^{2+}$ ($3d^7$, $l=3,\,s=3/2$). Moreover, despite the nominal $3+$ oxidation state of the magnetic ion in FeBO$_3$, the calculations predict a covalent, rather than ionic, character for the Fe-O bond: this results in an electronic configuration close to $3d^6$ ($l=2,\,s=2$), which is then expected to host a finite orbital moment.

In order to verify the theoretical predictions, we combined the well-established polarization dependence of NXMS \cite{blume_polarization_1988} with our novel rotating magnet technique \cite{dmitrienko_measuring_2014,Beutier2017}. The diffracted intensity arising from the long-range AFM order of the A(C,B)O$_3$ compounds was probed at several space-group forbidden reflections below the N\'{e}el transition temperature. A complete summary of the reflections measured for each compound can be found in the Supplemental Material \cite{supplemental_material} along with the relevant experimental parameters. The scattered signal is purely magnetic in origin, as proved by the fact that the signal vanishes upon warming above $T_N$ following the expected critical behaviour as a function of temperature \cite{supplemental_material}. The only exception is represented by the $(\bar{1}05)$ and $(\bar{2}07)$ reflections in CoCO$_3$, which will be discussed in \textbf{\S}\ref{sec:forbidden_scattering}. For each reflection, the signal was measured in both the $\sigma-\sigma'$ and $\sigma-\pi'$ channels as a function of a $360^{\circ}$ rotation of the external field in the basal plane of the crystal. The canted AFM structure rotates in response to the application of the field \cite{Beutier2017,dmitrienko_measuring_2014}: the corresponding magnetic field dependence of the scattered intensity can then be exploited to extract the relative orbital and spin contribution to the magnetic moment, as described in the following.

Given an incident $\sigma$-polarized x-ray beam, the NXMS amplitudes (neglecting a constant imaginary pre-factor) for $\sigma'$ and $\pi'$-polarized scattering read as follows\cite{blume_polarization_1988}:
\begin{align}
M_{\sigma\sigma'}&=\sin{2\theta}\,S_2 \nonumber\\
M_{\sigma\pi'}&=2\sin^2{\theta}\,[\cos{\theta}\,(L_1+S_1)+\sin{\theta}\,S_3]
\label{eq:scattering_amplitudes}
\end{align}
where $\theta$ is the Bragg angle of the measured $\text{(HKL)}$ reflection and $L_i$ and $S_i$ are the components of the orbital ($\mathbf{L}$) and spin ($\mathbf{S}$) structure factors in the $\mathbf{u}_1\mathbf{u}_2\mathbf{u}_3$ reference frame defined in Ref.~\onlinecite{blume_polarization_1988}, respectively. As shown in the schematic of Fig.~\ref{CoCO3_oscillations}, the latter is defined such that $\mathbf{u}_3$ is antiparallel to the scattering vector $\mathbf{Q}=\mathbf{k}_{out}-\mathbf{k}_{in}$, $\mathbf{u}_1$ lies in the scattering plane and points towards the detector and $\mathbf{u}_2$ is orthogonal to the scattering plane. The magnetic structure factors represent the Fourier transforms of the orbital and spin magnetization densities and thus directly depend on the direction of the magnetic moments. In the specific case of the magnetic reflections under study, they are given by (see Appendix~\ref{sec:structure_factors} for a detailed derivation):
\begin{align}
L_i &= C_L(\hat{\mu}_A^{(i)}-\hat{\mu}_B^{(i)})\nonumber\\ 
S_i &= C_S(\hat{\mu}_A^{(i)}-\hat{\mu}_B^{(i)})     
\label{eq:structure_factors}
\end{align}
where $\hat{\mu}_{A}^{(i)}$ [$\hat{\mu}_{B}^{(i)}$] is the $i$-th component of the magnetic moment (expressed as unit vector) of the A (B) sublattice along the $\mathbf{u}_i$ direction of the $\mathbf{u}_1\mathbf{u}_2\mathbf{u}_3$ frame. $C_L$ and $C_S$ are constants terms, whose ratio depends on the relative magnitude of the orbital ($\mathbf{l}$) and spin ($\mathbf{s}$) angular momenta and the orbital $[f_l(Q)]$ and spin $[f_s(Q)]$ form factors:
\begin{align}
\frac{C_L}{C_S}=\frac{l(Q)}{s(Q)}=\frac{\vert\mathbf{l}\vert}{\vert\mathbf{s}\vert}\frac{f_L(Q)}{f_S(Q)}
\label{eq:ratio}
\end{align}
where $Q$ is the modulus of the momentum transfer associated to the reflection $\text{(HKL)}$ considered and the form factors are defined such that $f_l(0)=f_s(0)=1$ (Appendix~\ref{sec:structure_factors}) [note that the quantity of Eq.~(\ref{eq:ratio}) could have been alternatively defined in terms of the magnetic moments $\bm{\mu}_L=-\mu_B\mathbf{l}$ and $\bm{\mu}_s=-2\mu_B\mathbf{s}$, leading to a factor of 2 difference].

\begin{figure}[htp]
	\centering
	\includegraphics[width=0.49\textwidth]{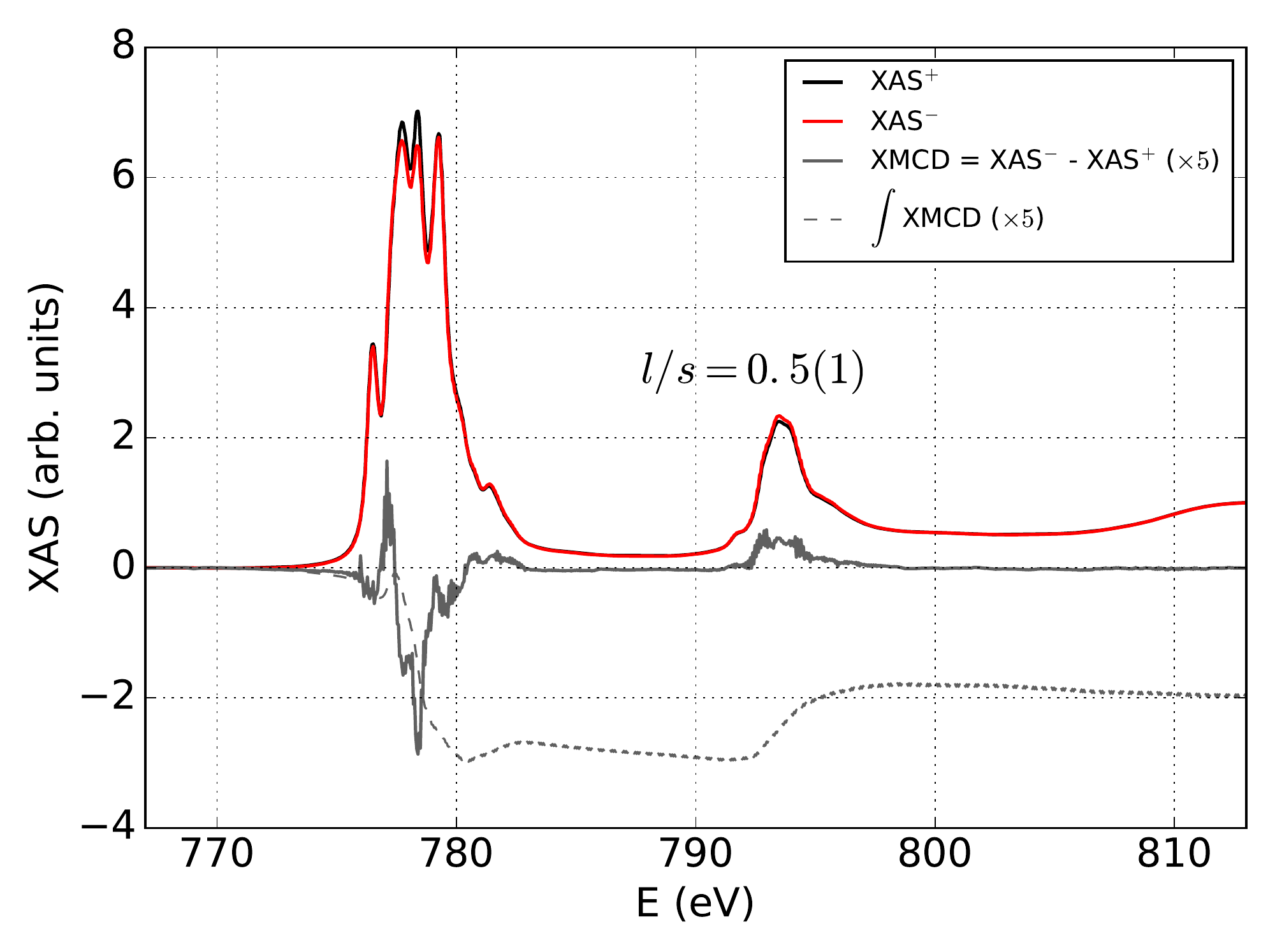}
	\caption[]{(Color online) Absorption spectra measured with the external magnetic field parallel (XAS$^-$) and antiparallel (XAS$^+$) to the helicity of the incident circularly polarized light and corresponding circular dichroism (XMCD). The data were collected at $T=3$~K using a magnetic field $\mu_0 H=0.4$~T applied in the $ab$ plane of the crystal. The dashed grey line represents the integrated XMCD signal used for the application of the sum rules: the $l/s$ value refers to corresponding orbital-to-spin angular momenta ratio. The XAS data are normalized such that the post-edge spectral weight is equal to unity.}
	\label{XMCD_low_field}
\end{figure}

In the case of negligible magnetocrystalline anisotropy (see \textbf{\S}~\ref{sec:anisotropy} for the case when this assumption no longer holds), the sum $\bm{\hat{\mu}}_A+\bm{\hat{\mu}}_B$ of the moments of the two sublattices aligns along the direction of the rotating external field $\mathbf{H}(\eta)$: the difference $\bm{\hat{\mu}}_A-\bm{\hat{\mu}}_B$ of Eq.~(\ref{eq:structure_factors}), perpendicular to the field, is forced to follow and causes the scattering amplitudes to vary accordingly.  After inserting Eq.~(\ref{eq:ratio}) and (\ref{eq:structure_factors}) into Eq.~(\ref{eq:scattering_amplitudes}), the corresponding diffracted intensities are described by the following relations:
\begin{align}
\label{eq:intensities}
I_{\sigma\sigma'}(\eta)&\propto\left|M_{\sigma\sigma'}(\eta)\right|^2=\displaystyle \left|\sin{2\theta}\,(\hat{\mu}_A^{(2)}-\hat{\mu}_B^{(2)})(\eta)\right|^2 \nonumber \\ 
I_{\sigma\pi'}(\eta)&\propto\left|M_{\sigma\pi'}(\eta)\right|^2=\displaystyle \left|2\sin^2{\theta}\,\left[\cos{\theta}\,\left(\frac{l(Q)}{s(Q)}+1\right)\cdot\right.\right. \nonumber\\
&\cdot\left.\left.(\hat{\mu}_A^{(1)}-\hat{\mu}_B^{(1)})(\eta)\right.\right.+
\left.\left.\sin{\theta}\,(\hat{\mu}_A^{(3)}-\hat{\mu}_B^{(3)})(\eta)\right]\right|^2
\end{align}
where the dependence of the magnetic moments differences $(\hat{\mu}_A^{(i)}-\hat{\mu}_B^{(i)})$ on the field angle $\eta$ has been emphasized.
The momentum-dependent orbital-to-spin ratio $l(Q)/s(Q)$ can be extracted through a fit to Eq.~(\ref{eq:intensities}) of the measured dependence of the diffracted intensities on the magnetic field direction in the $\sigma-\pi'$ and $\sigma-\sigma'$ polarization channels. Data for two different light polarizations are needed due to the arbitrary scale factor relating the modulus square of the scattering amplitudes to the measured intensities values. 

Representative data measured in CoCO$_3$ for four different magnetic reflections are displayed in Fig.~\ref{CoCO3_oscillations} along with the best fits to Eq.~(\ref{eq:intensities}). The data were collected by measuring the integrated intensity of the diffraction peak over a rocking scan of the sample at each value of the magnetic field angle $\eta$.  The magnetic intensity displays very well defined $180^\circ$-periodic sinusoidal oscillations, which are out of phase in the two polarization channels. For any given reflection, the $l(Q)/s(Q)$ ratio is encoded in the relative amplitude of the $\sigma-\sigma'$ and $\sigma-\pi'$ intensity modulations. It should be noted that the measured $\sigma-\sigma'$ / $\sigma-\pi'$ amplitude ratio is also subject to a trivial azimuth-dependent geometrical factor related to the components of the magnetic moments in the $\mathbf{u}_1\mathbf{u}_2\mathbf{u}_3$ reference frame [i.e. the quantities $(\mu_A^{(i)}-\mu_B^{(i)})$ of Eq.~(\ref{eq:intensities})]: this explains why the two symmetrically-equivalent $(107)$ and $(\bar{1}17)$ reflections show significantly different amplitudes in Fig.~\ref{CoCO3_oscillations} despite being characterized by the same value for the orbital-to-spin ratio. The data are of extremely high quality since the magnetic field measurements of Fig.~\ref{CoCO3_oscillations} are performed without moving the sample. Therefore, variations in the self-absorption and grain hopping caused by the sphere of confusion of the diffractometer, which affect more conventional azimuthal scans, are not present in this case.

\begin{figure}[htp]
	\centering
	\includegraphics[width=0.49\textwidth]{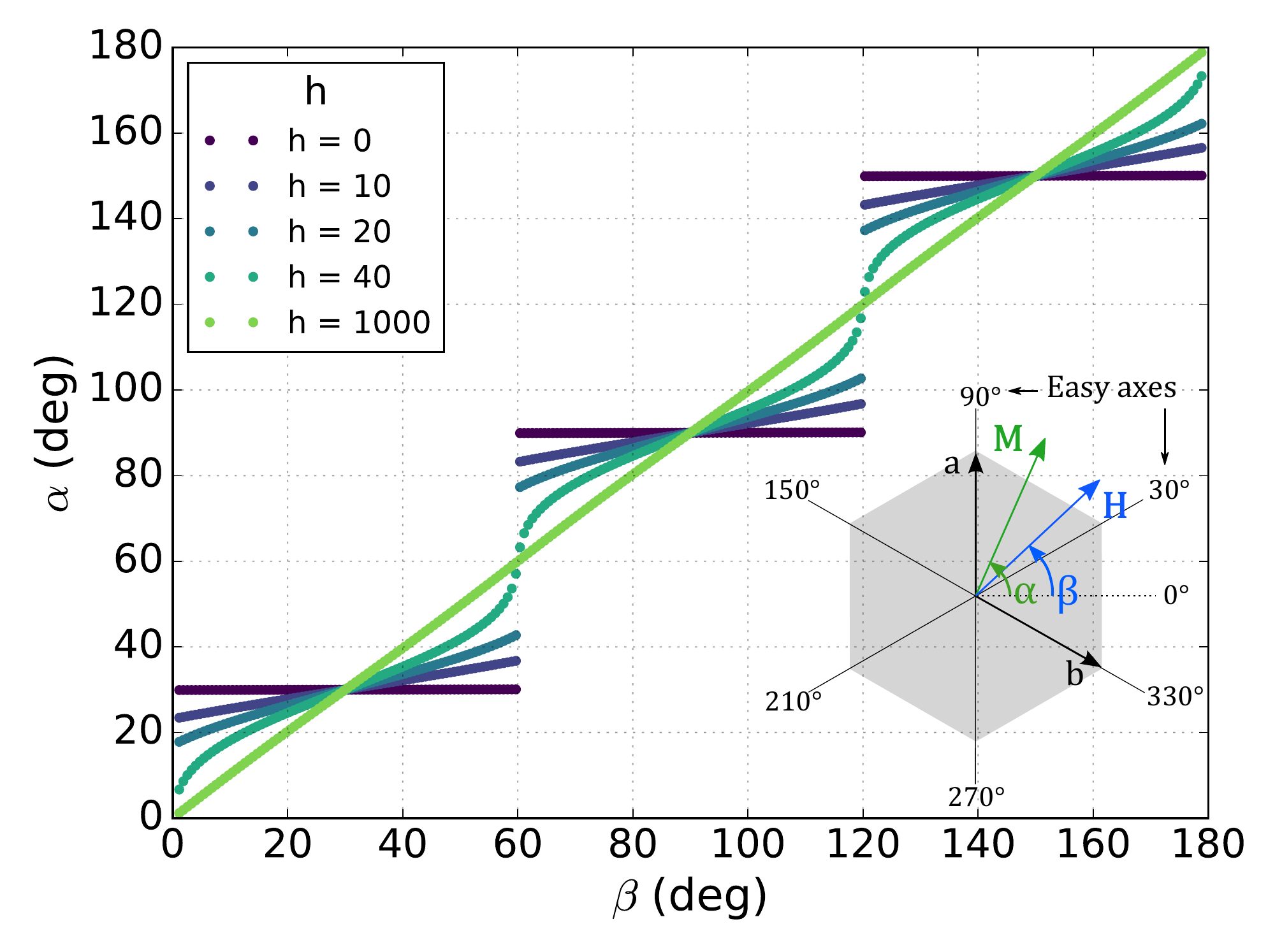}
	\caption[]{(Color online) Direction of the weak net magnetization $\mathbf{M}$ as a function of the external magnetic field $\mathbf{H}$ direction in the basal plane of the crystal for different values of the anisotropy parameter $h=\frac{\mu_0HM}{K}$ discussed in the text. Each curve $\alpha(\beta)$ corresponds to a solution of Eq.~(\ref{eq:hamiltonian_derivative}) for a different value of $h$. The inset shows the definitions of the magnetization ($\alpha$) and external magnetic field ($\beta$) angles in the  $ab$ basal plane of the  R$\bar{3}$c crystal structure.}
	\label{ANISOTROPY_magnetization_direction_2}
\end{figure}

\begin{figure*}[htp]
	\centering
	\includegraphics[width=\textwidth]{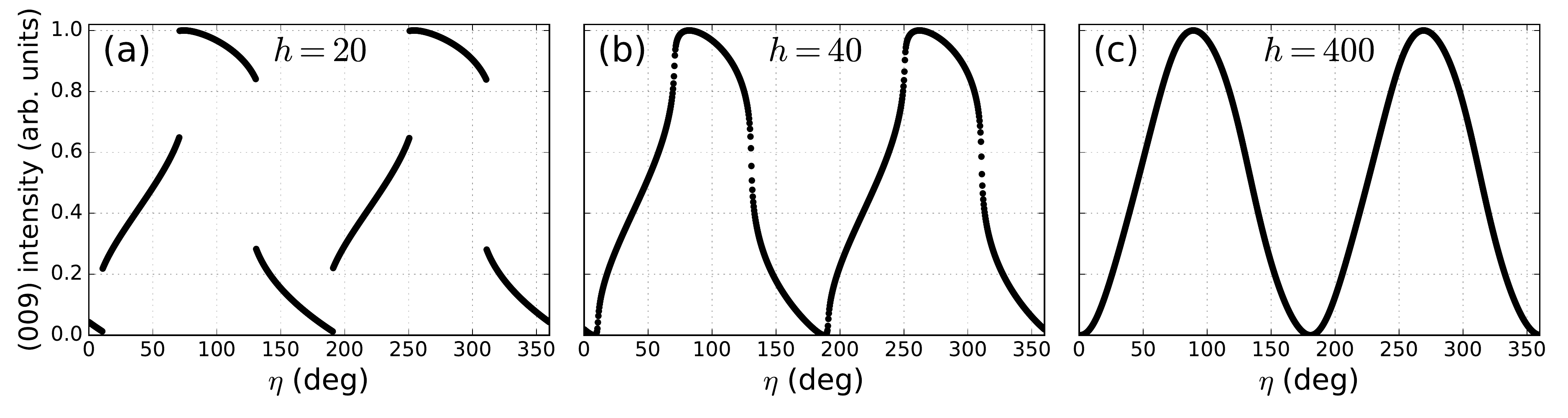}
	\caption[]{Calculated dependence of the $(009)$ magnetic reflection in the $\sigma-\pi'$ polarization channel on the external magnetic field direction for different values of the anisotropy parameter $h=\frac{\mu_0HM}{K}$ discussed in the text. The diffracted intensity was calculated using analogous solutions of Eq.~(\ref{eq:hamiltonian_derivative}) to the ones shown in Fig.~\ref{ANISOTROPY_magnetization_direction_2}.}
	\label{ANISOTROPY_simulated_intensity}
\end{figure*}

The $l(Q)/s(Q)$ values (averaged over all the measured temperatures and sample azimuths \cite{supplemental_material}) corresponding to different space-group forbidden reflections are shown as a function of the momentum transfer in Fig.~\ref{Orbital_moment} for the different compounds of the family. Following from Eq.~(\ref{eq:ratio}), the relative orbital and spin contribution to the total angular momentum of each compound, i.e. $\vert\mathbf{l}\vert/\vert\mathbf{s}\vert=l(0)/s(0)$, can be ultimately extracted by extrapolating the ratio $l(Q)/s(Q)$ to $Q=0$. This can be achieved by a fit to Eq.~(\ref{eq:ratio}) of the measured $l(Q)/s(Q)$ values assuming an isotropic approximation of the orbital and spin magnetic form factors (see Appendix~\ref{sec:structure_factors} for more details). The fit for the case of the CoCO$_3$ is reported as a dashed line in Fig.~\ref{Orbital_moment}. The resulting $\vert\mathbf{l}\vert/\vert\mathbf{s}\vert$ values are reported in Table~\ref{tab:calculations} along with the corresponding values from DFT calculations.

There is generally a very good agreement between the measurements and the calculations. Importantly, the trend across the series of compounds is confirmed: while MnCO$_3$ and FeBO$_3$ behave as almost pure spin systems, a significant unquenched orbital moment is found for both CoCO$_3$ and NiCO$_3$. In particular, the predicted large value of the orbital moment in CoCO$_3$ is confirmed. This is somewhat consistent with previous studies on crystals \cite{Neubeck2001,Radwanski2004, Lee2016} and thin films \cite{Csiszar2005} of CoO, where a large orbital moment was also found. As we will show in \textbf{\S}~\ref{sec:CoCO3}, the presence of a large orbital moment is confirmed by XMCD measurements and results in the emergence of several interesting phenomena in the physics of CoCO$_3$. When comparing the measurements with the calculations, it is important to bear in mind that due to the covalent bonding of the TM $3d$ orbitals with the oxygen $2p$ states, part of the magnetisation density appears on the ligand sites. While the NXMS measurements are sensitive to both, the DFT calculations neglect the oxygen contribution. This could explain, for instance, the partial discrepancy (still within the experimental uncertainty) between the measured and calculated value for CoCO$_3$.  This seems to be confirmed by the XMCD measurements outlined in \textbf{\S}~\ref{sec:CoCO3_XMCD}, which probes selectively the magnetization of the TM ion.

\begin{figure}[htp]
	\centering
	\includegraphics[width=0.49\textwidth]{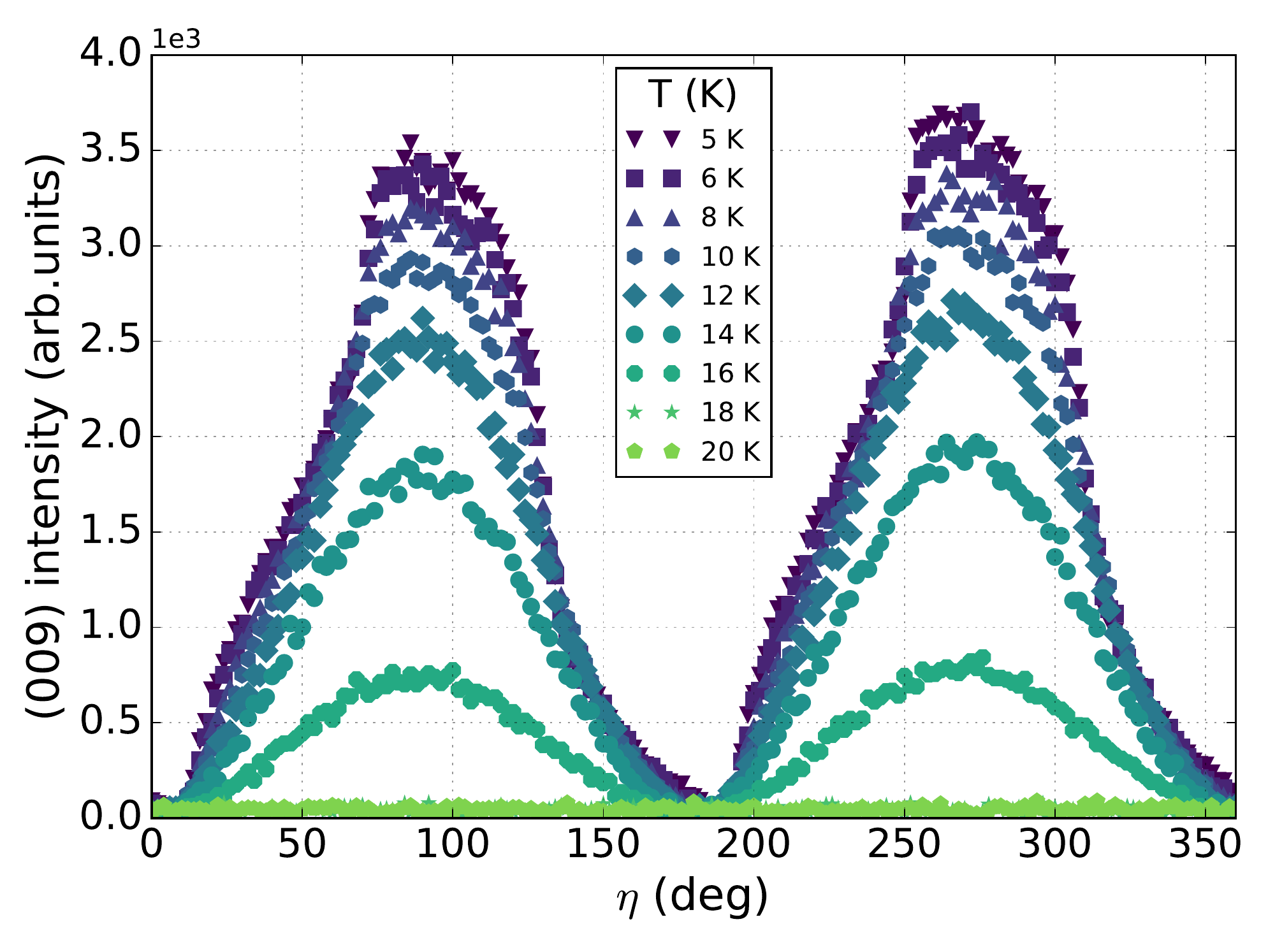}
	\caption[]{(Color online) Dependence of the CoCO$_3$ $(009)$ reflection on the magnetic field direction in $\sigma-\pi'$ at different temperatures across the N\'eel transition at $T_N=16.7(5)$~K \cite{supplemental_material}.}
	\label{ANISOTROPY_temperature_dependence}
\end{figure}

\section{The peculiar case of C\lowercase{o}CO$_3$}\label{sec:CoCO3}

\subsection{XMCD investigation of the orbital moment in CoCO$_3$}\label{sec:CoCO3_XMCD}

One of the main results of the previous section is the presence of an unusually large unquenched orbital moment in CoCO$_3$. In order to confirm this finding, we performed XMCD measurements at the Co $L$ edges as described in \textbf{\S}~\ref{sec:experimental_XMCD}. The results are shown in Fig.~\ref{XMCD_low_field}, where the absorption spectra obtained by combining the field and polarization reversal measurements are plotted along with the corresponding dichroism. The presence of a significant unquenched orbital moment is immediately evident from the much larger XMCD signal at the Co $L_3$  edge compared to $L_2$. The application of the sum rules for the spin ($\mu_s$) \cite{carra_x-ray_1993} and orbital ($\mu_l$) \cite{thole_x-ray_1992} magnetic moment to the integrated XMCD signal shown in Fig.~\ref{XMCD_low_field} leads to a value of the orbital-to-spin ratio $l/s=2\mu_l/\mu_s=0.5(1)$ which is  also confirmed by Co L-edge multiplet simulations ($l/s \approx 0.6$) as described in \textbf{\S}~\ref{sec:form_factor}. This value confirms the one derived from our NXMS measurements within the experimental uncertainty, thus further consolidating our findings. One could argue that the nominal value of the $l/s$ ratio found by means of XMCD is closer to the calculated one (Table~\ref{tab:calculations}), which neglects the oxygen contribution to the total magnetization density. This is perfectly consistent with the resonant nature of the absorption process which, contrary to NXMS, selectively probes the magnetization density localised on the Co$^{2+}$ ions.

\subsection{Single-ion anisotropy}\label{sec:anisotropy}

\begin{figure}[htp]
	\centering
	\includegraphics[width=0.49\textwidth]{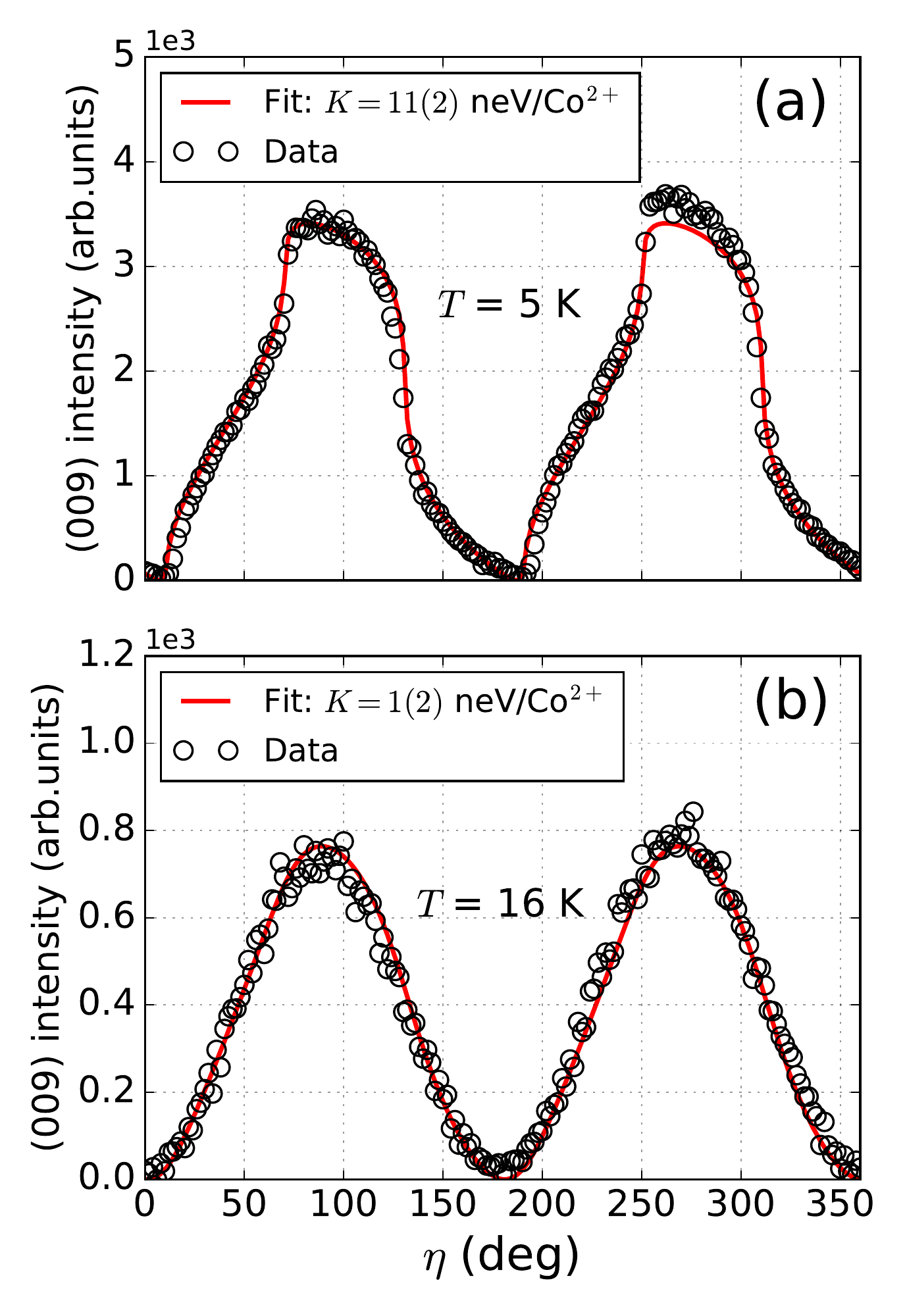}
	\caption[]{(Color online) Fit detail of two representative data sets of Fig.~\ref{ANISOTROPY_temperature_dependence} measured well below and close to the N\'eel transition. The solid lines represent the best fit to the calculated intensity (see Fig.~\ref{ANISOTROPY_simulated_intensity}) leaving the  anisotropy constant $K$ as a free parameter. The results of the fit of the data sets at all temperatures are shown in Fig.~\ref{ANISOTROPY_K_constant}.}
	\label{ANISOTROPY_fit}
\end{figure}

\begin{figure}[htp]
	\centering
	\includegraphics[width=0.49\textwidth]{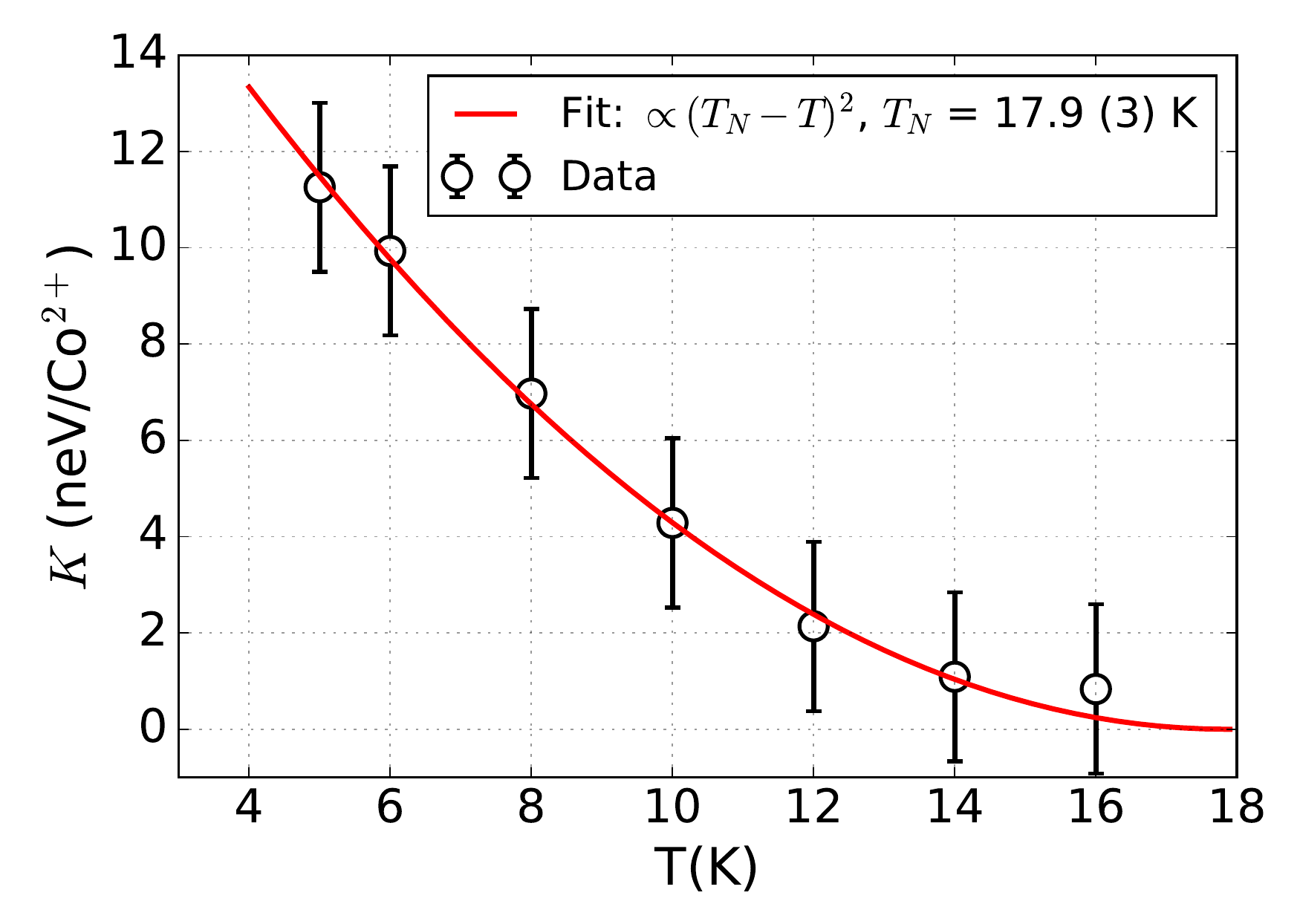}
	\caption[]{(Color online) Temperature dependence of the in-plane anisotropy constant $K$ in CoCO$_3$. The data points were obtained through analogous fits to the ones shown in Fig.~\ref{ANISOTROPY_fit}. The solid line represent the best fit to the quadratic law $\propto(T_N-T)^2$ proposed by \citet{Kaczer1963}.}
	\label{ANISOTROPY_K_constant}
\end{figure}

From the first-principles calculations, we also found that the large orbital moment in CoCO$_3$ strongly depends on the direction of the field. The latter is usually accompanied by a large magnetocrystalline anisotropy \cite{Bruno1989}, which we indeed observed and found to be about $9$~meV/Co$^{2+}$. Both effects were shown to originate from a peculiar combination of the crystal field and Coulomb correlations within the Co $d$-shell. In particular, the large magnetocrystalline anisotropy is caused by the pure $3z^2-r^2$-character of the lowest unoccupied orbital, which strongly favours in-plane orientation of the orbital moment, as demonstrated in detail in Ref.~\onlinecite{supplemental_material}. The magnetic anisotropy within the $\mathbf{ab}$ plane is expected to be significantly smaller. However, as we will show hereafter, its effect on the magnetic field dependence of the scattered intensity is clearly visible. For a crystal of space group R$\bar{3}$c the single-ion anisotropy in the $\mathbf{ab}$ plane is described by the following energy cost per unit volume \cite{Besser1967}:
\begin{align}
\left(\frac{E_{anis.}}{V}\right)=K_0(T)+K(T)\cos{6\alpha}
\label{eq:anisotropy}
\end{align}
where $K_0(T)$ and $K(T)$ are temperature-dependent constants (in energy per unit volume) which define the strength of the anisotropy and $\alpha$ is the angle describing the magnetization direction with respect to the crystal axes: this is defined such that the net magnetization resulting from the moments canting (which we shall refer to simply with the term ``magnetization'' from now on) is orthogonal to the $[100]$ crystallographic direction for $\alpha=0^{\circ}$ (see inset in Fig.~\ref{ANISOTROPY_magnetization_direction_2}). This 6-fold energy term is minimized for $\alpha=30^\circ+n60^\circ$ ($n$ integer index) and thus defines three main easy magnetization axes along the $[100]$, $[110]$ and $[0\bar{1}0]$ crystallographic directions. 

\begin{figure*}[htp]
	\centering
	\includegraphics[width=0.8\textwidth]{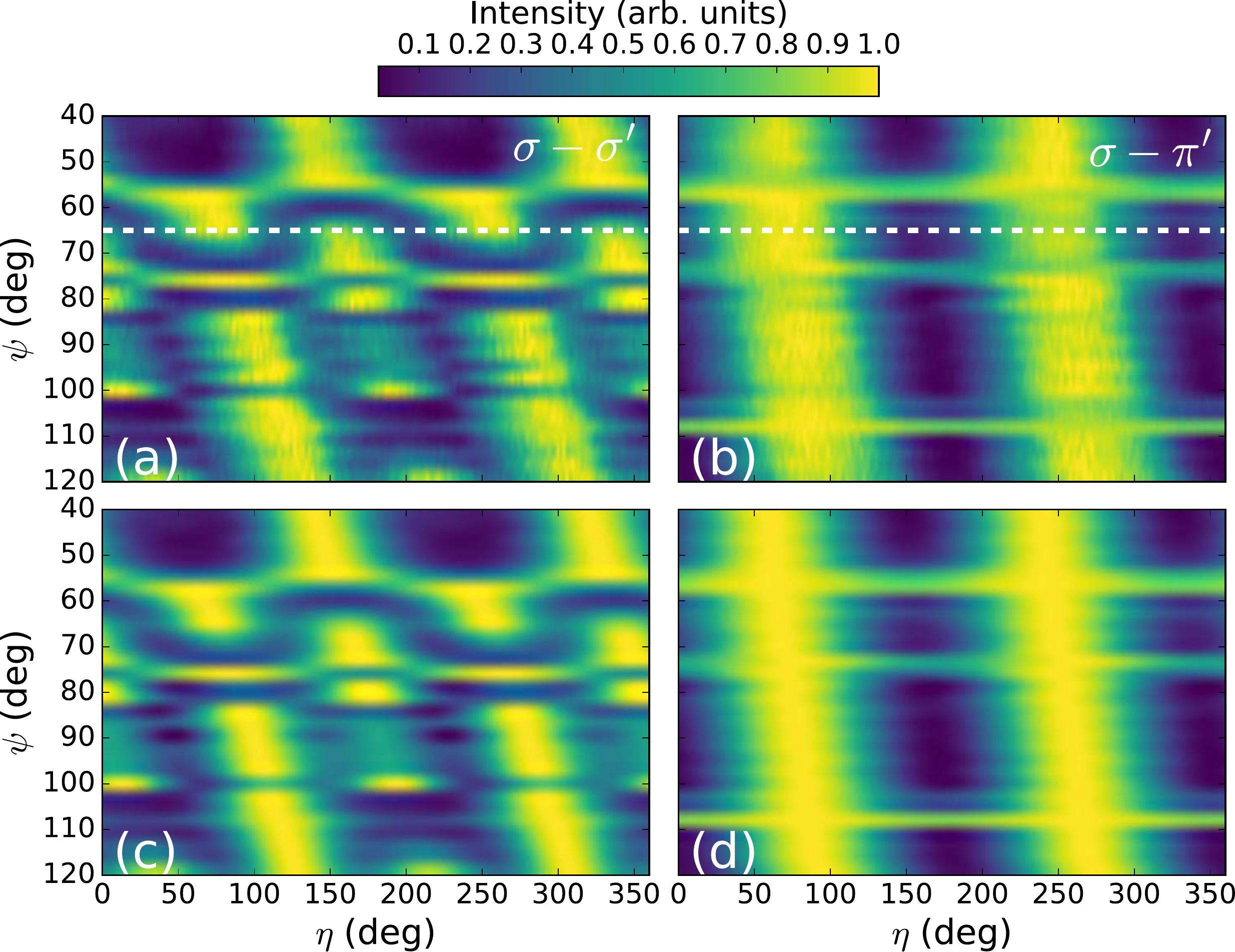}
	\caption[]{(Color online) (a-b) $(\bar{2}07)$ intensity magnetic field direction dependence as a function of the sample azimuth $\psi$ for two different polarization states of the diffracted x-ray beam. For each azimuth value and polarization channel, the  measured intensity (represented through the color scale) as a function of the field direction is normalized to its maximum value. A detail of the $\psi=65^\circ$ data sets (dashed horizontal lines) is shown in Fig.~\ref{CHARGE_fit_detail}. (c-d) Global fit of the $(\bar{2}07)$ magnetic field dependence shown in (a-b). The $\sigma-\sigma'$ and $\sigma-\pi'$ data at each azimuth value have been fitted to Eq.~(\ref{eq:intensity_forbidden}) and the second of Eq.~(\ref{eq:intensities}), respectively, as shown in Fig.~\ref{CHARGE_fit_detail} for $\psi=65^\circ$. The ratio between the magnetic and forbidden charge amplitudes in $\sigma-\sigma'$ was kept constant across all $\psi$ values, while the multiple scattering amplitude was left free to vary: the resulting values are plotted in Fig.~\ref{CHARGE_multiple_scattering}.}
	\label{CHARGE_colormap}
\end{figure*}

In the presence of an external magnetic field $\mathbf{H}(\beta)$ the total energy per unit volume can be written as [dropping the constant $K_0$ in Eq.~(\ref{eq:anisotropy})]:
\begin{align}
\frac{E}{V}&=\frac{E_{anis.}}{V}+\frac{E_{field}}{V}=\nonumber\\
&=K(T)\cos{6\alpha} - \mu_0HM(T)\cos{(\alpha-\beta)}
\label{eq:hamiltonian}
\end{align}
where $M(T)$ is the temperature-dependent magnitude of the magnetization. Analogous to the magnetization angle $\alpha$, $\beta$ defines the direction of the external magnetic field with respect to the orthogonal to the $[100]$ direction. This is related to the angle $\eta$ used to express the magnetic field dependence of the scattered intensity through the relation $\beta=\psi-\eta+60^{\circ}$, where $\psi$ is the sample azimuth and the $60^{\circ}$ offset is simply due to the initial magnet position with respect to the crystal axes. In Eq.~(\ref{eq:hamiltonian}), $E_{field}$ represents the Zeeman interaction of the net magnetic moment with the external field. For the case of negligible anisotropy considered in \textbf{\S}~\ref{sec:systematic_investigation} ($E_{anis.}\approx0$), the Zeeman term forces the magnetization to align parallel to the applied field ($\alpha=\beta$). In the general case of non-negligible anisotropy, however, $\mathbf{M}$ and $\mathbf{H}$ will lie along different directions. For any given direction $\beta$ of the external field, the equilibrium direction $\alpha$ of the magnetization is obtained by minimizing the Hamiltonian of Eq.~(\ref{eq:hamiltonian}):
\begin{align}
\label{eq:hamiltonian_derivative}
\frac{d}{d\alpha}\left(\frac{E}{KV}\right)=0
\end{align}

The solutions $\alpha(\beta)$ of Eq.~(\ref{eq:hamiltonian_derivative}) can be calculated numerically for different values of the dimensionless parameter $h(T)=\displaystyle\frac{\mu_0HM(T)}{K(T)}$, which expresses the relative strength of the Zeeman and anisotropy energy terms. These are plotted over a $180^\circ$-degree range of $\beta$ values in Fig.~\ref{ANISOTROPY_magnetization_direction_2}. The limiting case of negligible anisotropy considered in \textbf{\S}~\ref{sec:systematic_investigation} corresponds to large $h$ values and leads to the trivial solution $\alpha=\beta$. At the other extreme, for very large anisotropy values (small values of $h$) the magnetization is locked on the easy magnetization axes ($\alpha=30^\circ, 90^\circ\ldots$) regardless of the direction of the field and jumps discontinuously from one easy axes to another as the field rotates. A non-trivial 6-fold periodic function is obtained in the intermediate regime.  

The solutions $\alpha(\beta)$ can be used to calculate the magnetic structure factors~(\ref{eq:structure_factors}) and simulate the magnetic scattering intensities of Eq.~(\ref{eq:intensities}) for different $h$ values. The simulations are reported in Fig.~\ref{ANISOTROPY_simulated_intensity} for the $(009)$ $\sigma-\pi'$ intensity and three representative $h$ values. For negligible anisotropy (large $h$) a smooth sinusoidal oscillation is obtained, analogous to the data shown in Fig.~\ref{CoCO3_oscillations}. As the anisotropy increases ($h$ decreases) the intensity modulation takes on a peculiar ``shark-fin'' shape and eventually becomes discontinuous. Exactly the same trend is seen in the measured data as a function of temperature shown in Fig.~\ref{ANISOTROPY_temperature_dependence}. Increasing the temperature towards the N\'{e}el transition has, in this case, the effect of weakening the magnetocrystalline anisotropy: upon warming, the shark-fin shape progressively disappears and symmetric sinusoidal oscillations are recovered close to $T_N$.

\begin{figure}[htp]
	\centering
	\includegraphics[width=0.49\textwidth]{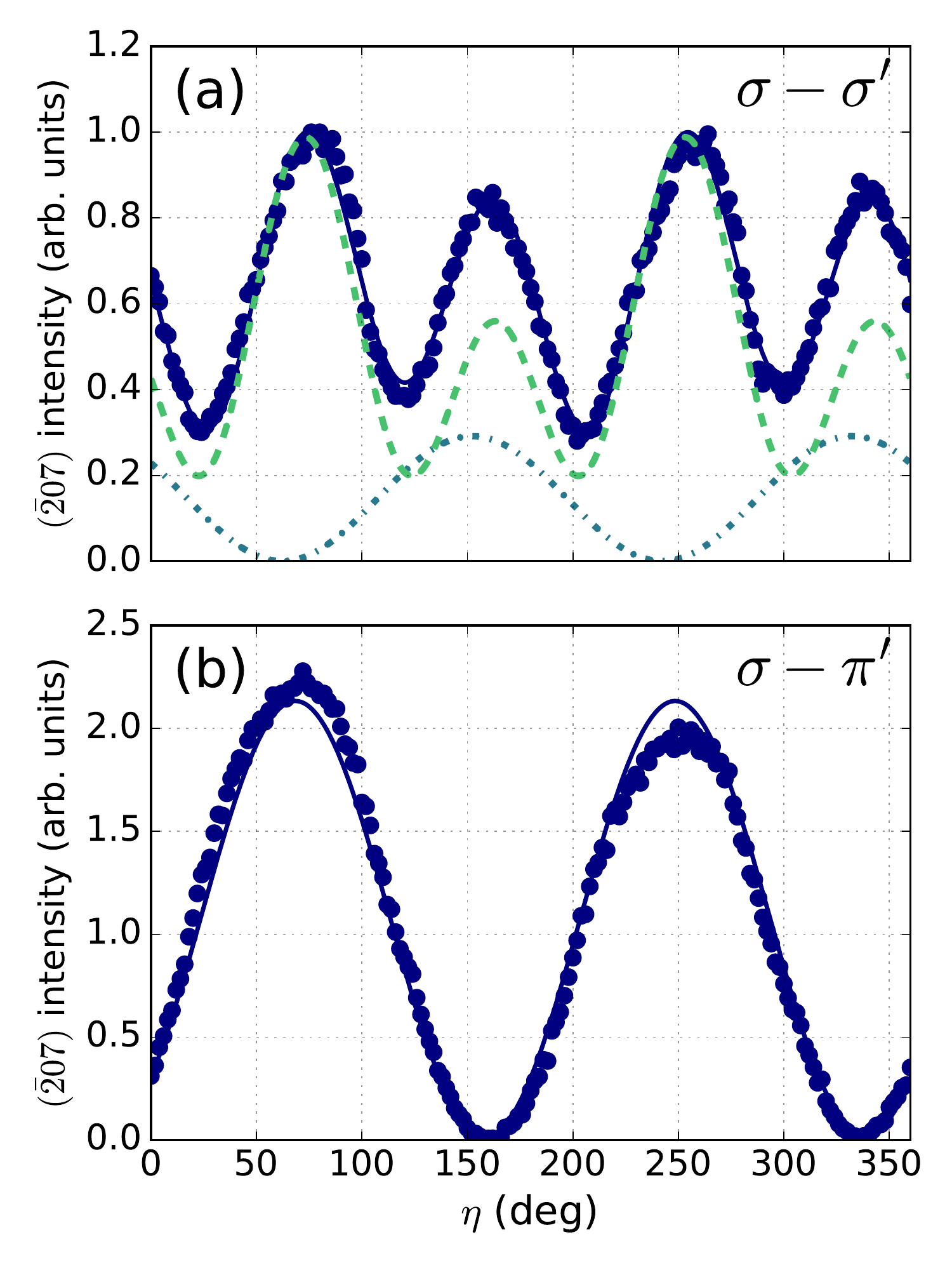}
	\caption[]{(Color online) $(\bar{2}07)$ intensity magnetic field dependence in (a) $\sigma-\sigma'$ and (b) $\sigma-\pi'$ at $\psi=65^{\circ}$ [dashed horizontal line of Fig.~\ref{CHARGE_colormap}(a-b)]. The symbols represent the measured diffracted intensity, while the solid line refer to the global fit as explained in the text. In (a) the dashed green line and the dash-dot light blue line correspond to the charge and magnetic contribution to the global fit, respectively. The data were collected at $T=4$~K and are normalized to the peak intensity of the $\sigma-\sigma'$ channel. A constant background originating from multiple scattering has been removed from the $\sigma-\pi'$ data set.}
	\label{CHARGE_fit_detail}
\end{figure}

The scattered intensity calculated from the solutions of Eq.~(\ref{eq:hamiltonian_derivative}) can be used to fit the experimental data of Fig.~\ref{ANISOTROPY_temperature_dependence} leaving the anisotropy constant $K(T)$ as a free fitting parameter and using the magnetization values of Ref.~\onlinecite{Meshcheryakov2007a} in the parameter $h$. The calculations reproduce extremely well the measurements, as shown in the two representative fits of Fig.~\ref{ANISOTROPY_fit} for data collected well below and close to the magnetic transition. The values of $K(T)$ obtained from the fits are displayed in Fig.~\ref{ANISOTROPY_K_constant}. As expected, the basal plane anisotropy constant decreases with increasing temperature: a quadratic dependence of the type $\propto(T_N-T)^2$ is observed over most of the temperature range explored, similar to what was reported by \citet{Kaczer1963}. We find $K=11(2)$~neV/Co$^{2+}$ at $T=4$~K, in agreement with a previous estimate\cite{Kaczer1963}. Although almost 6 orders of magnitude smaller than the out-of-plane value, the effect of a finite basal plane anisotropy is clearly visible in the data and proves the extremely high sensitivity of our novel rotating magnetic technique to small interaction terms of the magnetic Hamiltonian.

The anisotropy-induced distortion at low temperature is not as evident in the data of Fig.~\ref{CoCO3_oscillations} collected around $T=5$~K. The two sets of data were measured during different experiments on two different crystals and several factors might explain the observed discrepancy. As well as differences in the crystal quality (crystal defects could, for instance, impact the field dependence of the scattered intensity), a significant beam heating has been observed in several occasions and might have also played a role despite the precautions taken to minimize it. The latter is very sensitive to the exact experimental conditions (which were different for the two data sets), such as sample mounting and incident flux: a sample temperature just a few degrees higher than the nominal value could explain the apparent lack of anisotropy in Fig.~\ref{CoCO3_oscillations}. Moreover, unlike the data presented in Fig.~\ref{CoCO3_oscillations}, the measurements of Fig.~\ref{ANISOTROPY_temperature_dependence} were not collected integrating the intensity over a rocking scan at each value of the field angle. This allowed a much larger number of data points to be collected in a significantly shorter time (minutes compared to hours): both the coarser sampling and the averaging of any long term drifts could be at the origin of the apparent lack of any significant shark-fin distortion. It should also be noted that the magnetic field value depends on the exact position of the permanent magnet used for the measurements. This was fixed on the diffractometer rotational stages manually: a slightly different position between the measurements of Fig.~\ref{CoCO3_oscillations} and Fig.~\ref{ANISOTROPY_temperature_dependence} is likely to play a role in the observed discrepancy.

\subsection{Forbidden charge scattering}\label{sec:forbidden_scattering}

\subsubsection{Experimental data and empirical model}\label{sec:empirical_model}

\begin{figure}[htp]
	\centering
	\includegraphics[width=0.49\textwidth]{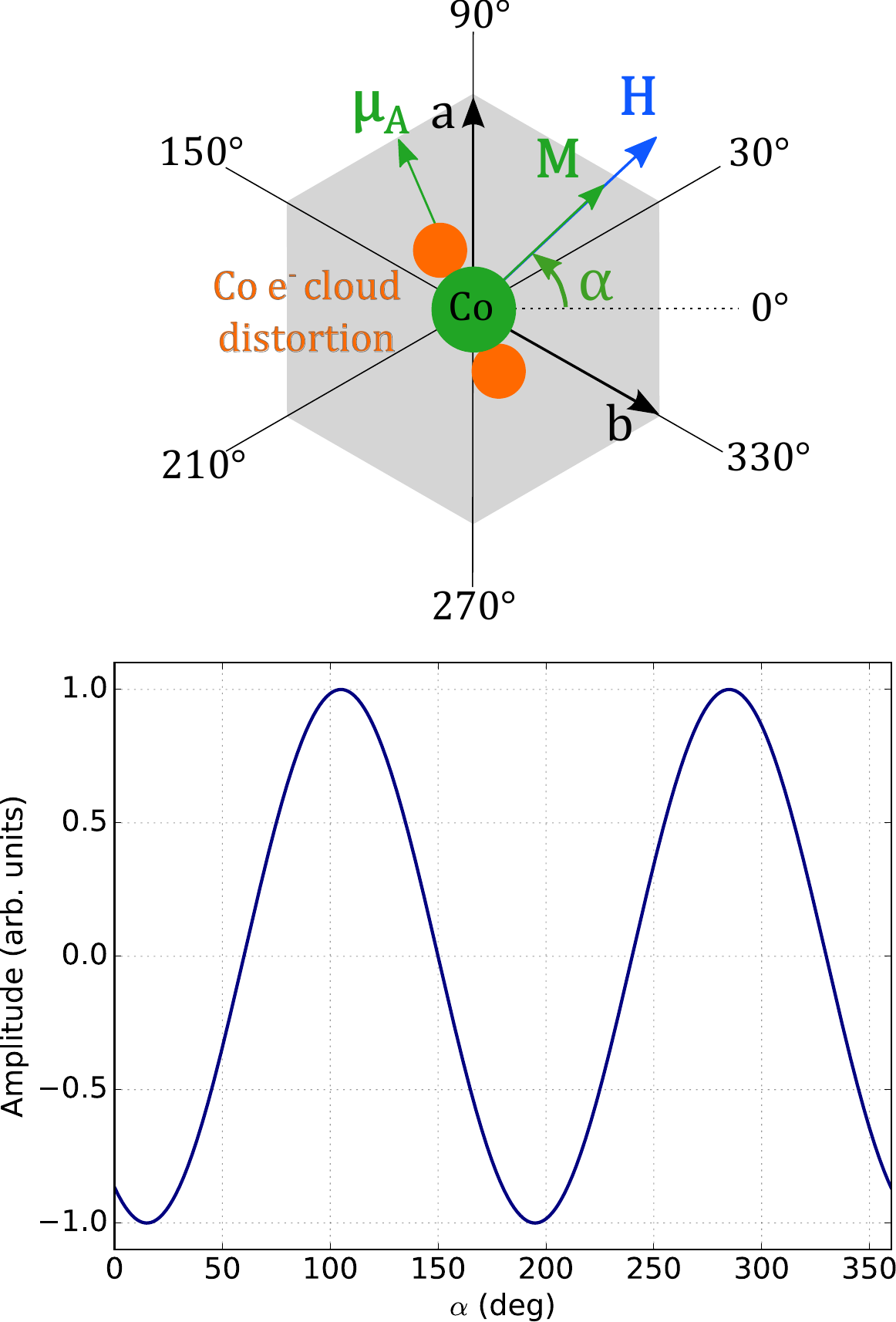}
	\caption[]{(Color online) Magnetic field dependence of the forbidden charge scattering amplitude normalized to its peak value. The drawing illustrates the empirical model for the Co electron cloud distortion discussed in the text. The Co ion of one of the two ferromagnetic sublattices (A) is shown together with the two negative charges used to model the elongation along the magnetic moment $\bm{\mu}_A$ direction. The definition of the field angle $\alpha$ is analogous to the one given in the inset of Fig.~\ref{ANISOTROPY_magnetization_direction_2} in the case of negligible magnetocrystalline anisotropy ($\mathbf{H}\parallel\mathbf{M}$).}
	\label{Empirical_model}
\end{figure}

\begin{figure}[htp]
	\centering
	\includegraphics[width=0.49\textwidth]{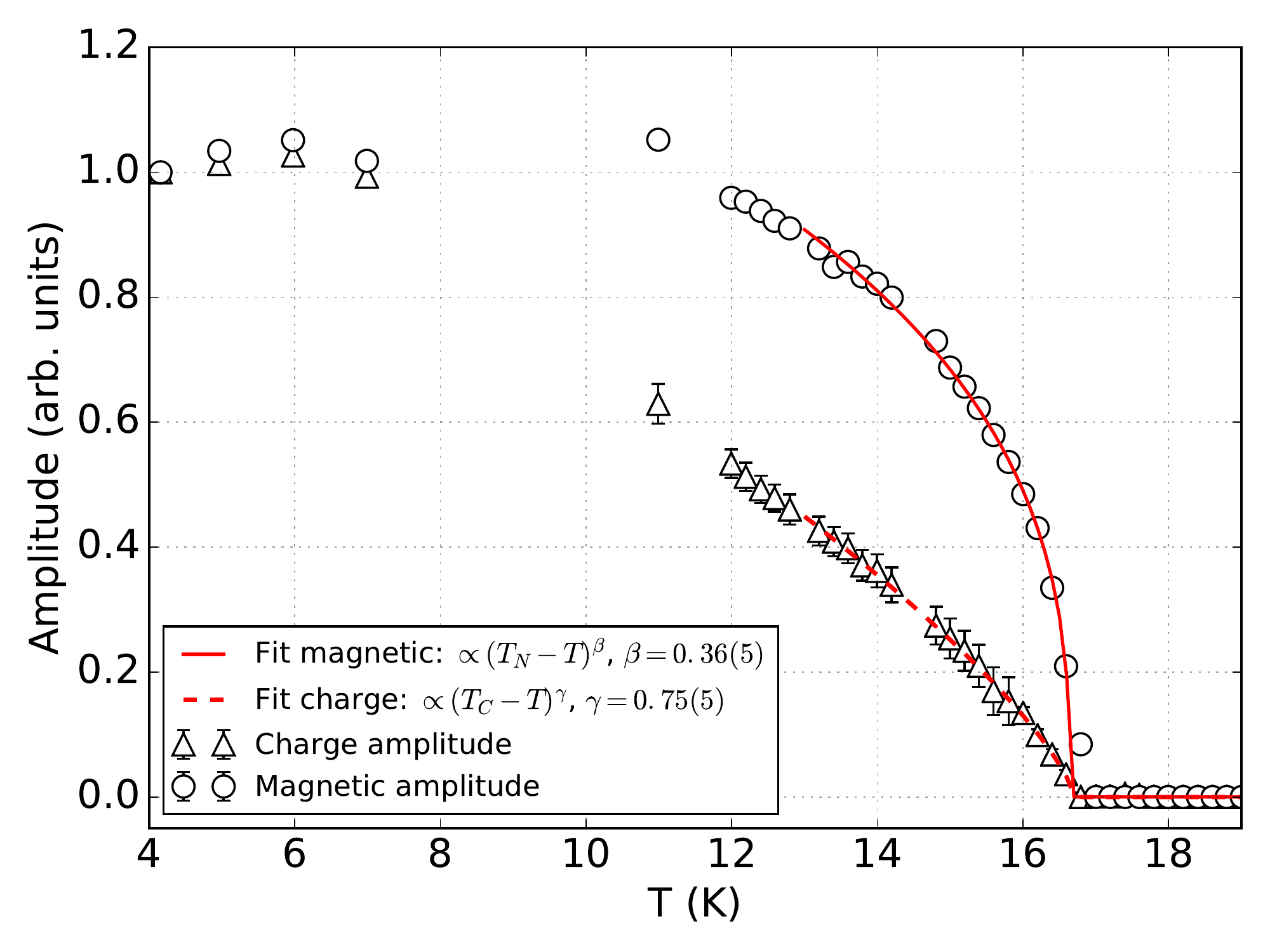}
	\caption[]{(Color online) Temperature dependence of the $(\bar{2}07)$ magnetic and charge amplitude for $\psi=65^\circ$. The data points were obtained through fits analogous to the one of Fig.~\ref{CHARGE_fit_detail}. The solid and dashed lines refer to the best fit to a power law for the magnetic and charge amplitude with critical exponents $\beta$ and $\gamma$, respectively.}
	\label{CHARGE_temperature_dependence_fit}
\end{figure}

One of the most striking manifestations of the large orbital moment in CoCO$_3$ consists in the appearance of a peculiar interference pattern in the magnetic field dependence of space-group forbidden reflections. As argued hereafter, this is the result of the presence of a subtle, extremely weak, contribution to the scattered intensity induced by the ordered magnetic moment below the N\'{e}el transition. The same effect has not been observed in the other compounds of the series \cite{supplemental_material} and is thus to be considered a distinctive aspect of the physics of CoCO$_3$. Of all the space-group forbidden reflections that we have measured, this interference is clearly visible only for the $(\bar{2}07)$ and $(\bar{1}05)$. Here, as shown in Fig.~\ref{CHARGE_colormap}(a) for the $(\bar{2}07)$ reflection, the dependence of the scattered intensity in the $\sigma-\sigma'$ channel on the magnetic field direction displays abrupt variations with the sample azimuth. A similar effect is also seen by varying the energy of the incident x-rays\cite{supplemental_material}. This contrasts the scattered intensity in $\sigma-\pi'$ [Fig.~\ref{CHARGE_colormap}(b)], which, regardless of the $\psi$ and energy value, exhibits the normal 2-fold oscillation seen for the other reflections (Fig~\ref{CoCO3_oscillations}). 

A detail of the $(\bar{2}07)$ magnetic field dependence at $\psi=65^\circ$ [dashed lines in Fig.~\ref{CHARGE_colormap}(a-b)] is shown in Fig.~\ref{CHARGE_fit_detail}. Magnetic scattering alone cannot account for the nearly 4-fold pattern observed in the measured intensity in $\sigma-\sigma'$: an extra contribution, displayed as a green dashed line in Fig.~\ref{CHARGE_fit_detail}(a), must be introduced. The latter is, in turn, the result of two interfering scattering amplitudes of charge origin: (i) a sinusoidally-oscillating forbidden scattering term, $C_{\sigma\sigma'}^{FS}(\eta)$, plotted in Fig.~\ref{Empirical_model} as a function of the field direction and (ii) a field-independent multiple-scattering amplitude, $C_{\sigma\sigma'}^{MS}$. As we will argue more extensively later on, both $C_{\sigma\sigma'}^{FS}(\eta)$ and $C_{\sigma\sigma'}^{MS}$ are real and, as a result, do not interfere with the magnetic amplitude, which is purely imaginary (i.e. a $\pi/2$ phase shift is present between the charge and magnetic contributions). The total scattered intensity in the $\sigma-\sigma'$ channel is therefore given by:
\begin{align}
I_{\sigma\sigma'}^{(\bar{2}07),(\bar{1}05)}(\eta)&=I_{\sigma\sigma'}^{Magnetic}+I_{\sigma\sigma'}^{Charge}\propto \nonumber\\
&\propto\left|M_{\sigma\sigma'}(\eta)\right|^2+\left\vert C_{\sigma\sigma'}^{FS}(\eta)+C_{\sigma\sigma'}^{MS}\right\vert^2
\label{eq:intensity_forbidden}
\end{align}
where $I_{\sigma\sigma'}^{Magnetic}=\left|M_{\sigma\sigma'}(\eta)\right|^2$ is given by the first line of  Eq.~(\ref{eq:intensities}). The charge origin of $C_{\sigma\sigma'}^{FS}(\eta)$ is suggested by the absence of the interference effect in the rotated polarization channel and further confirmed by its temperature dependence (Fig.~\ref{CHARGE_temperature_dependence_fit}): the latter exhibits a critical exponent twice as large as the magnetic one, as expected for magnetic-induced charge scattering \cite{Lovesey1996}.

The forbidden amplitude stems from a peculiar distortion of the Co$^{2+}$ electron cloud induced by the magnetic moment in the magnetically ordered phase. While a microscopic description requires detailed calculations of the Co$^{2+}$ ground-state wave function (see \textbf{\S}~\ref{sec:form_factor}), most aspects of the resulting scattering process are captured by the simple ``toy model'' sketched in the drawing of Fig.~\ref{Empirical_model}. This assumes a small elongation of the Co$^{2+}$ electron cloud along the magnetic moment ($\bm{\mu}$) direction, which is modelled by artificially adding a pair of negative charges to either side of each Co$^{2+}$ ion along $\bm{\mu}$. The electron cloud distortion reduces the symmetry of the crystal in the magnetically ordered phase such that, for an arbitrary field direction, only the inversion centre is left [space group $\text{P}\overline{1}$ (No.2)]. The two extra charges are set to rigidly follow the rotation of $\bm{\mu}$ as this is dragged around by the external field, thus originating the field-dependent term $C_{\sigma\sigma'}^{FS}(\eta)$ shown in Fig.~\ref{Empirical_model}. The latter interferes with the multiple scattering amplitude and gives rise to the observed magnetic field dependence.

\begin{figure}[htp]
	\centering
	\includegraphics[width=0.49\textwidth]{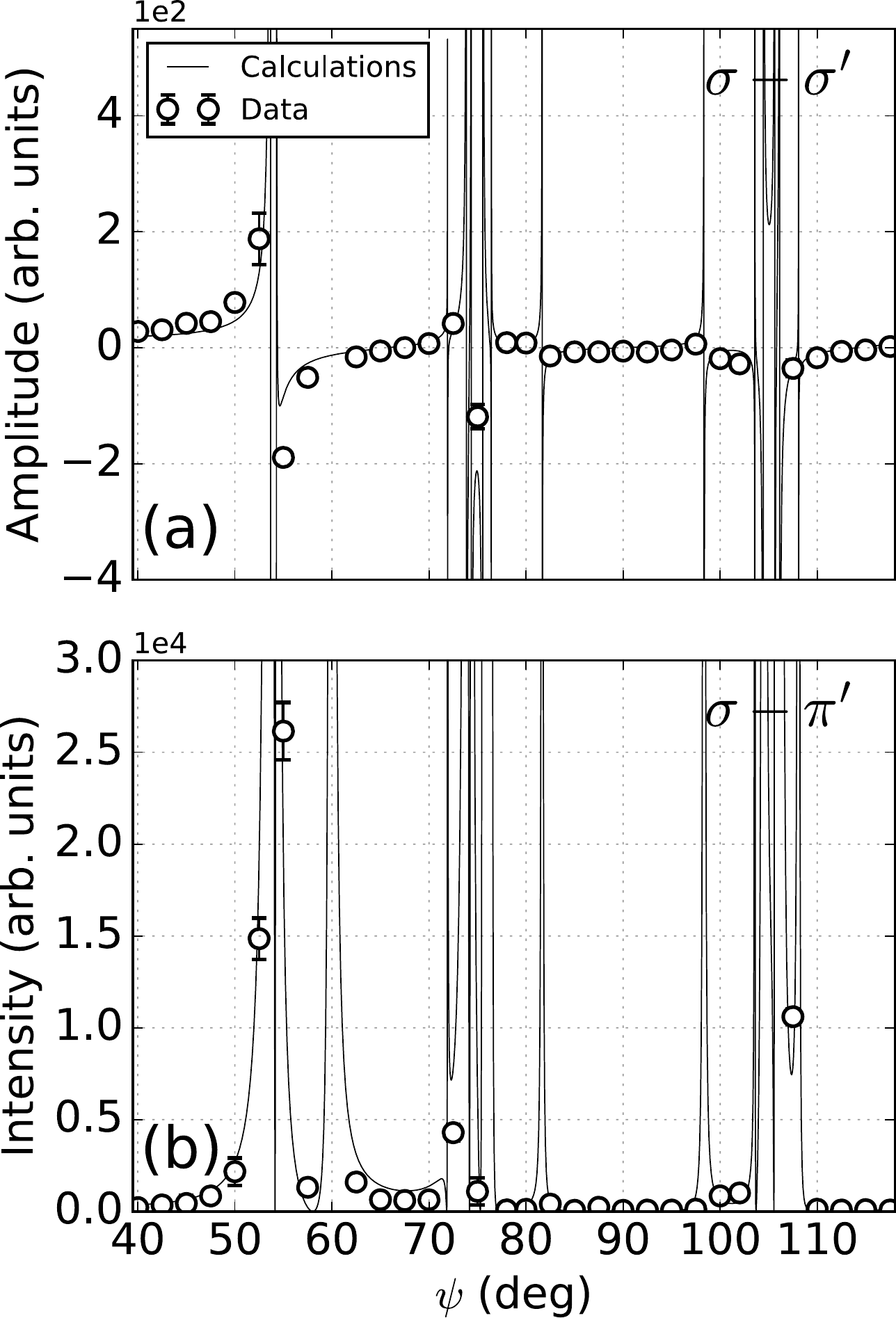}
	\caption[]{Multiple scattering (a) $\sigma-\sigma'$ amplitude and (b) $\sigma-\pi'$ intensity azimuthal dependence. The data points are the results of the fit of the data shown in Fig.~\ref{CHARGE_colormap}(a-b), while the solid line represents calculations performed using a mixed kinematical/dynamical approach where the standard kinematical structure factors of the secondary and tertiary reflection are weighted by terms calculated in a dynamical framework.}
	\label{CHARGE_multiple_scattering}
\end{figure}

The multiple scattering amplitude $C_{\sigma\sigma'}^{MS}$ plays a key role in our observations. In particular, it explains the azimuthal dependence of the scattered intensity of Fig.~\ref{CHARGE_colormap}(a). Contrary to standard Bragg diffraction (also referred to as two-wave diffraction), where the diffracted radiation originates from a single scattering event of the primary beam, in multiple-wave diffraction the secondary beam originated from the scattering of the incident x-rays can act as a primary beam for a second scattering process, thus giving rise to a tertiary reflection \cite{Weckert1997}. This results in additional diffraction peaks, which can appear at nominally forbidden $\text{(HKL)}$ values. Although much weaker than Bragg reflections, multiple scattering peaks can have a comparable intensity to the magnetic ones. The condition for generating multiple-wave diffraction is much more stringent than in the two-wave case: as a result, the multiple scattering amplitude displays a strong dependence on the sample azimuth (see Fig.~\ref{CHARGE_multiple_scattering}) as opposed to Bragg scattering, which does not depend on $\psi$. Moreover, while the latter does not change the polarization of the primary beam, multiple scattering can in general give rise to both $\sigma'$ and $\pi'$-polarized radiation.


The $\psi$ values at which multiple scattering occurs can be calculated \cite{Nisbet2015a,supplemental_material}, and thus avoided, following a simple kinematical approach. Nonetheless, broad tails are also present away from the nominal scattering condition and generally result in a residual contribution in both polarization channels. Because of the inversion centre of the R$\bar{3}$c space group, all structure factors, including the multiple scattering one, are real if one considers only Thomson scattering far from any absorption edge. Therefore, multiple scattering interferes with the ``forbidden'' amplitude (which turns out to be of Thomson nature and is thus real), but not with non-resonant magnetic scattering, which is out of phase by $90^{\circ}$, and gives rise to the dramatic evolution with the sample azimuth reported in Fig.~\ref{CHARGE_colormap}. In the absence of the forbidden amplitude, as is the case for the $\sigma-\pi'$ intensity of all reflections and the $\sigma-\sigma'$ one when the additional term $C_{\sigma\sigma'}^{FS}(\eta)$ is absent (Fig.~\ref{CoCO3_oscillations}), multiple scattering simply results in a constant background superimposed to the intensity of magnetic origin. A significant multiple scattering background is responsible for the high-intensity streaks visible in the $\sigma-\pi'$ color map of Fig.~\ref{CHARGE_colormap}(b).

\bgroup
\footnotesize
\def\arraystretch{1.5}
\begin{table*}[htp]
	\centering
	
	\resizebox{\textwidth}{!}{
		\begin{tabular}{cccccc|ccccccccccc}
			\hline\hline
			\multicolumn{6}{c|}{Hamiltonian parameters (eV)}                                          & \multicolumn{10}{c}{Expectation values}                                                                                                                             \\ \hline
			$F_{3d-3d}^{(2)}$ & $F_{3d-3d}^{(4)}$ & $10D_q$ & $D_\sigma $ & $H_{ex}$ & SOC & $\hat{H}$ & $\hat{S}^2$ & $\hat{L}^2$ & $\hat{J}^2$ & $\hat{S}_{x}$ & $\hat{S}_{y}$ & $\hat{S}_{z}$ & $\hat{L}_{x}$ & $\hat{L}_{y}$ & $\hat{L}_{z}$ \\ \hline
			7.9072           & 5.0463            & 1       & 0.06        & 0.0018  &  0.052   & -2.780   & 3.745       & 11.675      & 22.278      & -0.791 &  0.051 & 0.000  & -0.501  & 0.032 & 0.000        \\ \hline\hline
		\end{tabular}
	}
	\caption{Hamiltonian parameters used to perform the multiplet calculations of the Co$^{2+}$ ground-state wave function in CoCO$_3$ and corresponding expectations values of relevant quantities. The spin and orbital angular momenta components of one of the two magnetic sublattices, in units of $\hbar$, are expressed in the local $\mathbf{xyz}$ cubic frame of the CoO$_6$ octahedra, such that $\mathbf{x}$ is parallel to the $\mathbf{a}$  crystallographic axis and $\mathbf{z}$ is parallel to the crystallographic $\mathbf{c}$ axis.}
	
	\label{tab:multiplet}
\end{table*}
\egroup

\begin{figure*}[htp]
	\centering
	\includegraphics[width=0.8\textwidth]{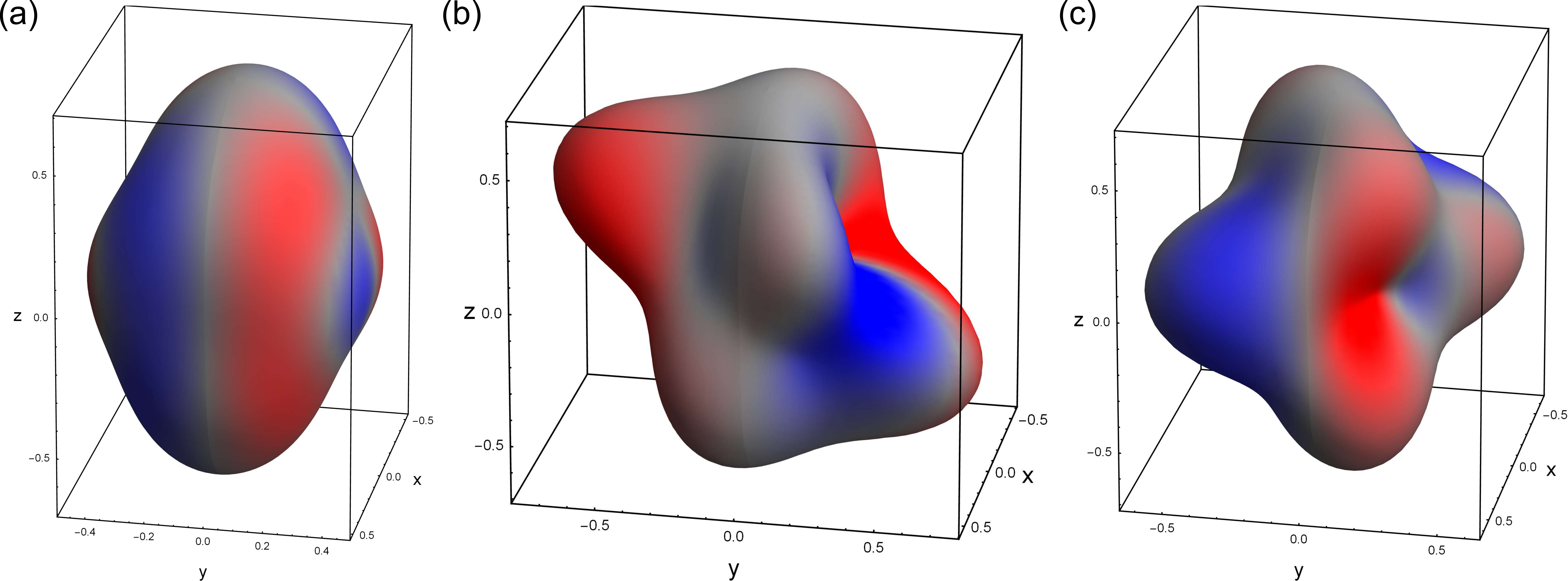}
	\caption[]{(Color online) Charge density of the Co$^{2+}$ $3d$ valence electrons for an external field direction (a) $0^{\circ}$, (b) $60^{\circ}$ and (c) $90^{\circ}$ away from the crystallographic $\mathbf{a}$ axis as derived from the multiplet calculations discussed in the text. The plots refer to one of the two Co$^{2+}$ clusters used for the calculations (cluster two). The colours correspond to different spin directions with red and blue for down and up character, respectively.}
	\label{CHARGE_charge_density}
\end{figure*}

The interference between the amplitude induced by the electron cloud distortion and multiple scattering reproduces extremely well our data: this is clearly shown by the color map of Fig.~\ref{CHARGE_colormap}{c}, which displays the fit of the measured intensity of Fig.~\ref{CHARGE_colormap}(a) to Eq.~(\ref{eq:intensity_forbidden}). The fit of the $\sigma-\pi'$ intensity [Fig.~\ref{CHARGE_colormap}(d)] was performed using the second line of Eq.~(\ref{eq:intensities}), analogous to \textbf{\S}~\ref{sec:systematic_investigation}. An arbitrary positive scale factor, constant throughout all $\psi$ values, was used for the forbidden charge amplitude of Fig.~\ref{Empirical_model} in the fit of the $\sigma-\sigma'$ intensity. This is because the empirical model is not capable of reproducing a physically-meaningful value of the scattering amplitude. On the other hand, the phase of the oscillations (including its sign) is correctly predicted. This is elegantly proved by the values of the multiple scattering amplitude $C_{\sigma\sigma'}^{MS}$ extracted from the fits of Fig.~\ref{CHARGE_colormap}(c-d), which are reported in Fig.~\ref{CHARGE_multiple_scattering} along with the amplitude calculated using a  mixed kinematical/dynamical approach where the standard kinematical structure factors of the secondary and tertiary reflection were weighted by terms calculated in a dynamical framework \cite{Kokubun1998,Kokubun2004}. The $\psi$ dependence of the measured amplitude, in particular its sign, is remarkably consistent with the calculations, thus confirming the correctness of the forbidden amplitude phase.

The empirical model also grasps other significant features of the forbidden amplitude that are confirmed by our form factor calculations (\textbf{\S}~\ref{sec:form_factor}). In particular, it predicts a vanishing amplitude when (i) the canting of the Co$^{2+}$ magnetic moments is set to 0 (perfect AFM alignment) or (ii) a specular $\text{(00L)}$ reflection is considered. Within this simple model, a non-vanishing forbidden amplitude is expected to be present also for the two equivalent reflections $(107)$, $(\bar{1}17)$. However, this term appears to be much smaller than the magnetic contribution \cite{supplemental_material} and is not clearly visible in the measured data. The latter are well described by the magnetic scattering cross sections alone. It should be noted that we also investigated alternative models based on displacements of the different atomic species inside the unit cell: however, a satisfactory description of the observed forbidden amplitude could not be achieved.


\subsubsection{Microscopic model and role of SOC}\label{sec:form_factor}

In order to achieve a microscopic understanding of the electron cloud distortion induced by the ordered moment, we derived the $3d$ electron ground-state wave function for different directions of the external field by means of multiplet calculations. The latter were carried out using a Hartree-Fock method in the mean-field approximation, using the code RCN \cite{Cowan1967} for the radial part of the Co$^{2+}$ wave function and Quanty \cite{Haverkort2012} for the angular part.  The ground state was computed separately for two clusters of Co$^{2+}$ ions: the latter correspond to the two unique orientations of CoO$_6$ octahedra of the R$\bar{3}$c crystal structure, one rotated by $164^\circ$ about the $\mathbf{c}$ axis with respect to the other \cite{supplemental_material}. The $[\bar{1}11]$ direction of the local octahedral frame is parallel to the unit cell $\mathbf{c}$ axis for both clusters; the $[110]$ direction of cluster one (two) is rotated by $22^{\circ}$ ($-142^\circ$) about $\mathbf{c}$ relative to the unit cell $\mathbf{a}$ axis \cite{supplemental_material}. The Hamiltonian used for the calculations consists of the following terms: (i) Coulomb interaction ($F_{3d-3d}^{(2)}$, $F_{3d-3d}^{(4)}$), (ii) crystal field ($10D_q$, $D_\sigma$), (iii) SOC, (iv) magnetic exchange ($H_{ex}$) and (v) Zeeman term of interaction with the external field. $H_{ex}$ is a mean-field term which mimics the effect of the field produced by the ordered moments. The values of the main parameters used for the calculations are summarized in Table~\ref{tab:multiplet}: these were obtained refining the initial atomic values to reproduce the XMCD spectra of Fig.~\ref{XMCD_low_field} (see Supplemental Material\cite{supplemental_material}). The corresponding ground-state Hamiltonian expectation values are also summarized in Table~\ref{tab:multiplet}: despite the absolute values of the spin and orbital moments are somewhat different from the ones reported in Table~\ref{tab:calculations} \cite{supplemental_material}, a large value of the orbital contribution ($l/s\approx0.6$) is confirmed. Moreover, as well as reproducing the absorption spectra, the electronic structure of the Co$^{2+}$ ion thus calculated is consistent with the one presented in Ref.~\onlinecite{meshcheryakov_crystal_2004} and reproduces the experimental XMCD spectra well\cite{supplemental_material}. Further information on the crystal field parameters and additional details on the calculations are reported in the Supplemental Material \cite{supplemental_material}.

The results of the calculations confirm that, as predicted by our empirical model, the charge density of the valence $3d$ electrons depends on the magnetic moment orientation. This is shown in Fig.~\ref{CHARGE_charge_density}, where a real-space representation of the charge density is reported for different directions of the external field. The Co$^{2+}$ ground-state wave functions for different field directions can then be used to calculate the corresponding atomic form factor: the resulting $(\bar{2}07)$ and $(\bar{1}05)$ scattering amplitudes show a sinusoidal magnetic field dependence analogous to the one of Fig.~\ref{Empirical_model}. Consistent with the empirical model, the amplitude vanishes for specular $\text{(00L)}$ reflections and when the magnetic moment canting angle is set to 0. Most importantly, the multiplet calculations show that the amplitude also vanishes when the SOC is artificially switched off\cite{supplemental_material}. This attributes the magnetic-moment-induced distortion of the Co$^{2+}$ electron cloud to the coupling between lattice and magnetic degrees of freedom driven by SOC and further highlights the fundamental role played by the large unquenched orbital moment in the physics of CoCO$_3$.

\begin{figure}[htp]
	\centering
	\includegraphics[width=0.49\textwidth]{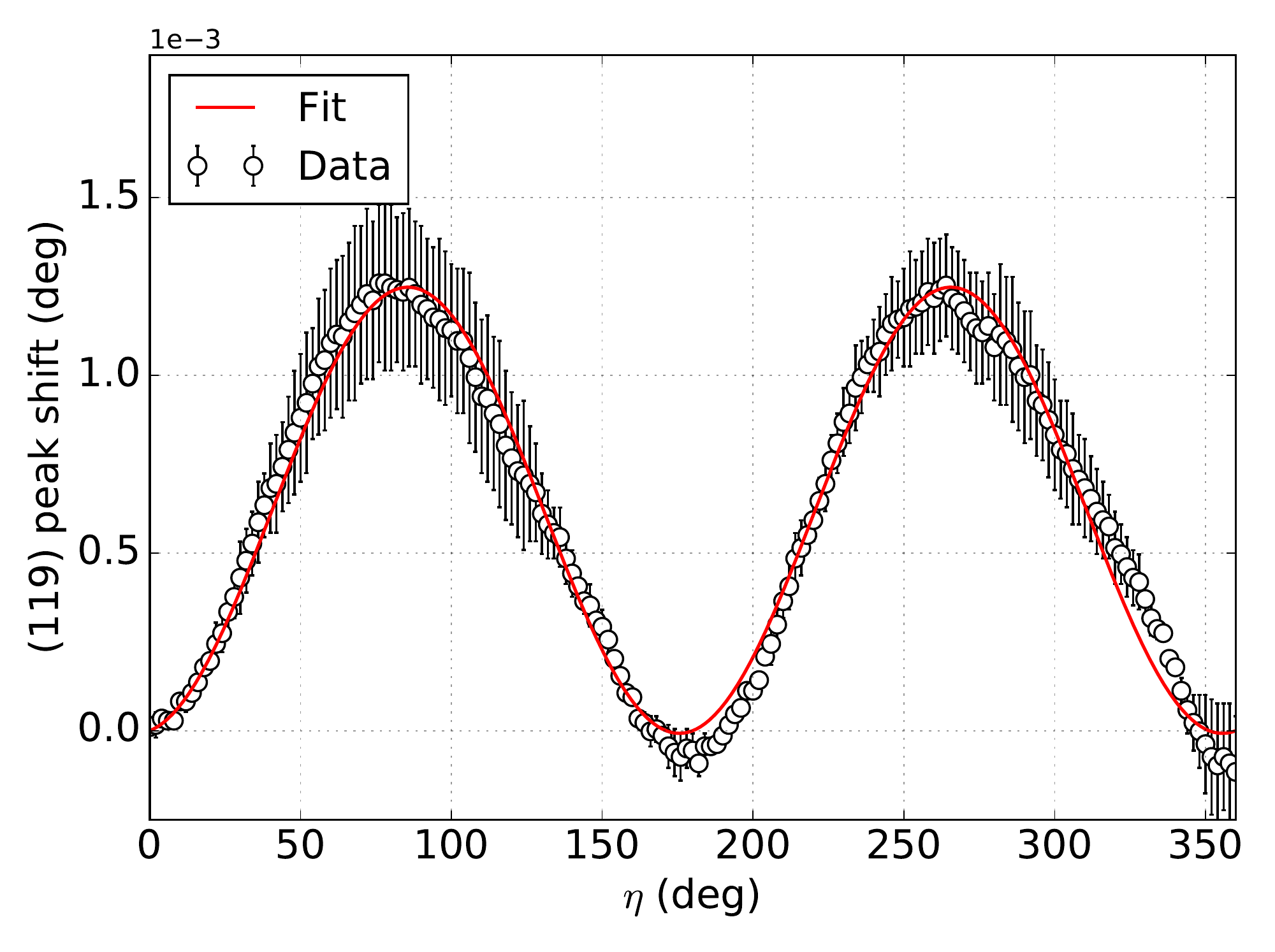}
	\caption[]{(Color online) Angular shift of the $(119)$ Bragg peak as a function of the magnetic field direction at $T=5$~K. The solid line represents a fit to the shift calculated using the unit cell deformation of Eq.~(\ref{eq:unit_cell_deformation}). The deformation of the in-plane lattice parameters resulting from the fit is shown in Fig.~\ref{CHARGE_(119)_deformation}.}
	\label{CHARGE_(119)_peak_shift}
\end{figure}

\subsection{Magneto-striction}\label{sec:magnetostriction}

\begin{figure}[htp]
	\centering
	\includegraphics[width=0.49\textwidth]{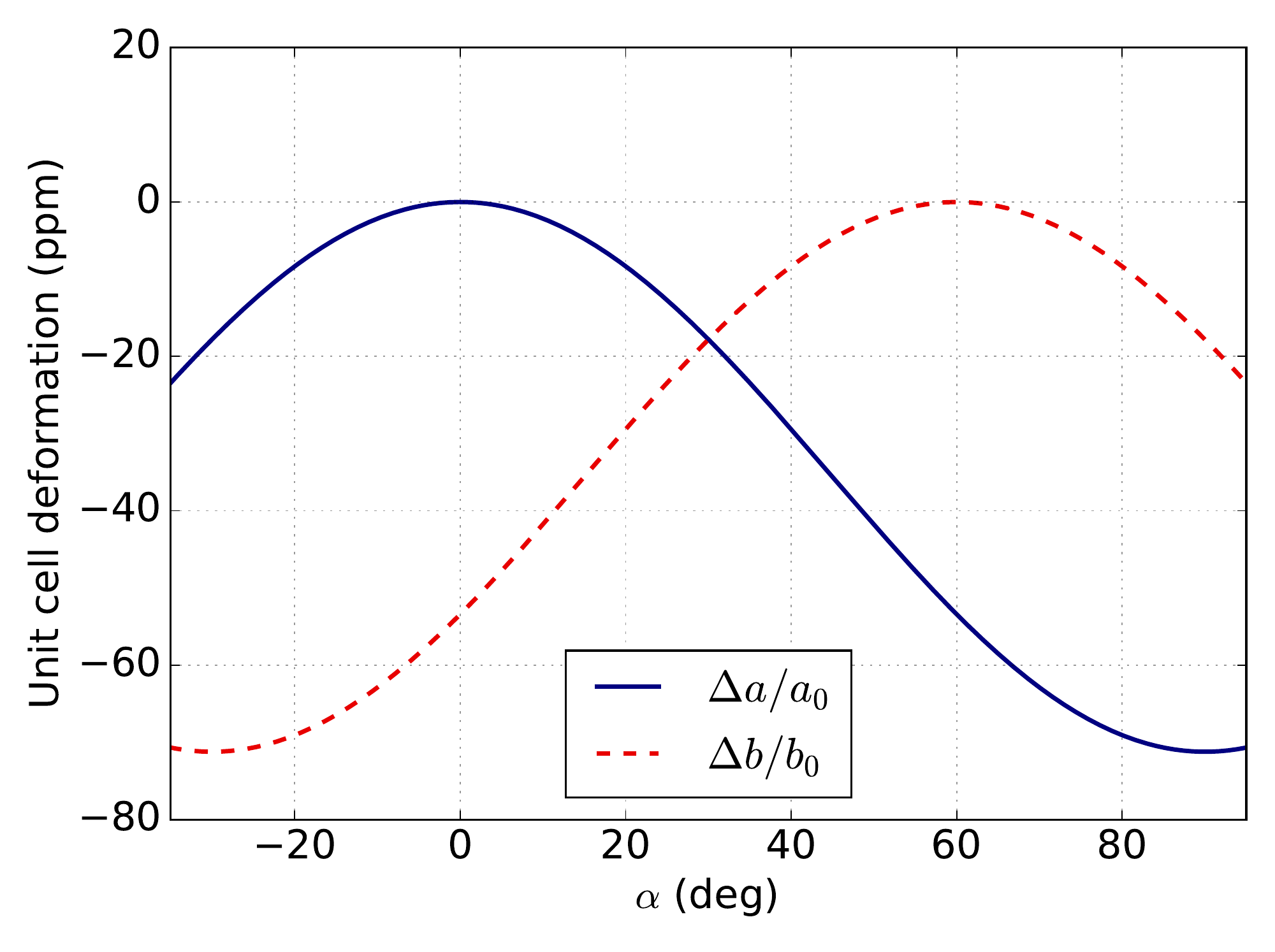}
	\caption[]{(Color online) Relative deformation of the in-plane lattice parameters of the R$\bar{3}$c hexagonal unit cell at $T=5$~K as obtained from the fit of the $(119)$ angular shift of Fig.~\ref{CHARGE_(119)_peak_shift}.}
	\label{CHARGE_(119)_deformation}
\end{figure}

Another distinctive evidence of the elongation of the Co$^{2+}$ electron cloud along the magnetic moment direction is the resulting expansion of the unit cell in-plane lattice parameters. This is revealed by the angular shift of the Bragg peak of symmetry-allowed reflections as a function of the field direction in the magnetically-ordered phase (Fig.~\ref{CHARGE_(119)_peak_shift}). Given an expansion $\Delta l>0$ of the unit cell along $\bm{\mu}$, the Bragg angle $\theta$ magnetic field dependence is correctly described by the following field-dependent lattice parameters distortion:
\begin{align}
a(\alpha)&=a_0-\Delta l\sin^2{\alpha} \nonumber\\
b(\alpha)&=b_0-\Delta l\sin^2({60^\circ-\alpha})
\label{eq:unit_cell_deformation}
\end{align}
where $a_0$ ($b_0$) is the value of the lattice parameter $a$ ($b$) when the magnetic field is orthogonal to the $\mathbf{a}$ ($\mathbf{b}$) axis [$\alpha=0^\circ$ ($60^\circ$)].
The magnitude of the unit cell distortion $\Delta l$ can be obtained by fitting the measured shift of the Bragg peak to the one calculated through the lattice parameters of Eq.~(\ref{eq:unit_cell_deformation}). This is shown by the red solid line of Fig.~\ref{CHARGE_(119)_peak_shift}. The resulting unit cell deformation along the $\mathbf{a}$ and $\mathbf{b}$ axes is plotted as function of the field direction in Fig.~\ref{CHARGE_(119)_deformation}. The deformation amounts to $\approx70$~ppm at $T=5$~K (which correspond to a change of $\approx35$~fm in the lattice parameters) and decreases upon warming towards $T_N$ following the same critical behaviour of magnetic scattering \cite{supplemental_material}. This further confirms the magnetostrictive origin of the Bragg peak oscillations of Fig.~\ref{CHARGE_(119)_peak_shift} and constitutes further evidence of the coupling between crystallographic and magnetic properties induced by the large unquenched orbital moment.

\section{Concluding remarks}\label{sec:conclusions}

In conclusion, our combined DFT, NXMS and XMCD investigation of a series of isostructural weak ferromagnets led to the following findings:
\begin{itemize}
	\item A non-trivial evolution of the orbital contribution to the magnetic moment with the filling of the TM ion $3d$ orbitals is present across the series. In particular, the value of the orbital moment was found to be particularly large for CoCO$_3$ as confirmed by both NXMS and XMCD.
	\item In CoCO$_3$, SOC couples the large orbital moment and the spin of the Co$^{2+}$ ion and results in a strong single-ion uniaxial anisotropy and a much smaller, although still clearly visible in our NXMS data, basal plane anisotropy.
	\item SOC is also responsible for a distortion of the Co$^{2+}$ $3d$ electron cloud in the magnetically order phase: the latter is evidenced by a sizeable magneto-striction and, more spectacularly, by the appearance of a forbidden scattering amplitude at space-group forbidden reflections.
\end{itemize}
Our results combined together highlight the importance of SOC in the physics of weak ferromagnets and show how, even in the case of $3d$ transition metal oxides, SOC can have a significant impact on the magnetic properties of the system whenever the orbital degrees of freedom are not quenched. Finally, our investigation also proves the ability of modern first-principles calculations to predict the properties of materials which exhibit magnetoelectric coupling, Skyrmion lattices and other non-collinear magnetic ordering.

\begin{acknowledgments}
The authors acknowledge Diamond Light Source for time on beamline I16 under Grants No. MT-12479, MT-13608 and MT-16227 and beamline I10 under Grants No. SI-12478 and SI-15987 and the XMaS facility at the European Synchrotron
Radiation Facility for beamtime under Grant No. BM-28-01-
966 (data in Supplemental Material\cite{supplemental_material}). The authors would also like to thank P. Bencok and the other I10 beam line's staff members for the support provided during the XMCD and bulk magnetization measurements. The work of V.V.M. is supported by the Russian Science Foundation Grant 18-12-00185. A. I. L. acknowledges the support of Grant No. DFG SFB-668 and the excellence cluster CUI. M. I. K. acknowledges support from ERC Advanced Grant No. 338957 FEMTO/NANO. Y. O. K. acknowledges the computational resources provided by the Swedish National Infrastructure for Computing (SNIC) and Uppsala Multidisciplinary Center for Advanced Computational Science (UPPMAX). Y.O.K. and V.V.M. acknowledge the financial support provided by The Swedish Foundation for International Cooperation in Research and Higher Education (STINT). The work of V.E.D. was supported partly by the grant ‘NANO’ of the Presidium of Russian Academy of Sciences and partly by the Ministry of Science and Higher Education within the State assignment FSRC ``Crystallography and Photonics'' RAS.
\end{acknowledgments}

\appendix

\section{Magnetic structure factors}\label{sec:structure_factors}

Following from \citet{blume_polarization_1988}, the spin and orbital magnetic structure factors which appear in the magnetic scattering amplitudes of Eq.~(\ref{eq:scattering_amplitudes}) are defined as follows:
\begin{align}
\mathbf{S}&=\displaystyle\sum_{i} \mathbf{s}_i f_s^i(\mathbf{Q}) e^{i\mathbf{Q}\cdot\mathbf{r}_i} \nonumber\\
\mathbf{L}&=\displaystyle\sum_{i} \mathbf{l}_i f_l^i(\mathbf{Q}) e^{i\mathbf{Q}\cdot\mathbf{r}_i}
\label{eq:structure_factors_1}
\end{align}
where $\mathbf{s}_i$ ($\mathbf{l}$) and $f_s^i(\mathbf{Q})$ [$f_l^i(\mathbf{Q})$] is the spin (orbital) angular momentum and magnetic form factor of the $i$-th magnetic ion in the crystal unit cell. $\mathbf{r}_i=u\mathbf{a}+v\mathbf{b}+w\mathbf{c}$ is the corresponding position vector (with $\mathbf{a},\,\mathbf{b},\,\mathbf{c}$ direct lattice basis vectors) and $\mathbf{Q}=\text{H}\mathbf{a}^{\,*}+\text{K}\mathbf{b}^{\,*}+\text{L}\mathbf{c}^{\,*}$ is the momentum transfer of the chosen $\text{(HKL)}$ reflection (with $\mathbf{a}^{\,*},\,\mathbf{b}^{\,*},\,\mathbf{c}^{\,*}$ reciprocal lattice basis vectors). The summation runs over the magnetic ions of the magnetic unit cell, which, in the case of the compounds of interest for the present paper, coincides with the crystallographic one.
In the case where $\mathbf{j}=\mathbf{s}+\mathbf{l}$ is a good quantum number, the following relations hold between the spin and orbital angular momenta and the total magnetic moment $\bm{\mu}$ of the magnetic ion \cite{Lovesey1996}:
\begin{align}
\label{eq:good_quantum_number}
\mathbf{s}&=\displaystyle(g-1)\mathbf{j}=\frac{1-g}{g}\frac{1}{\mu_B}\bm{\mu} \nonumber\\ 
\mathbf{l}&=\displaystyle(2-g)\mathbf{j}=\frac{g-2}{g}\frac{1}{\mu_B}\bm{\mu}
\end{align}
where $g$ is the Land\'{e} factor.
Inserting Eq.~(\ref{eq:good_quantum_number}) into Eq.~(\ref{eq:structure_factors_1}) and extending the summation of Eq.~(\ref{eq:structure_factors_1}) to the TM ions inside the R$\bar{3}$c hexagonal unit cell, the structure factors can be expressed as follows:
\begin{align}
\label{eq:structure_factor_2}
\mathbf{S}=\frac{1-g}{g}f_s(\mathbf{Q})\frac{\mu}{\mu_B}\sum_{i}{\hat{\mu}_i}e^{i\mathbf{Q}\cdot\mathbf{r}_i}=C_S(\bm{\hat{\mu}}_A-\bm{\hat{\mu}}_B)   \nonumber \\    
\mathbf{L}=\frac{g-2}{g}f_l(\mathbf{Q})\frac{\mu}{\mu_B}\sum_{i}{\hat{\mu}_i}e^{i\mathbf{Q}\cdot\mathbf{r}_i}=C_L(\bm{\hat{\mu}}_A-\bm{\hat{\mu}}_B)
\end{align}
Here, $\hat{\mu}=\bm{\mu}/\vert\bm{\mu}\vert$ is the magnetic moment unit vector, $\displaystyle C_S=3\,\frac{g-1}{g}f_s(\mathbf{Q})\frac{\mu}{\mu_B}$ and $\displaystyle C_L=3\,\frac{2-g}{g}f_l(\mathbf{Q})\frac{\mu}{\mu_B}$. The quantity $3(\bm{\hat{\mu}}_A-\bm{\hat{\mu}}_B)$ simply comes from computing the summation of Eq.~(\ref{eq:structure_factor_2}) for the six TM ions (Wyckoff site $b$ with multiplicity 6) of the $R\bar{3}$c hexagonal cell. Finally, Eq.~(\ref{eq:structure_factors}) is obtained directly from Eq.~(\ref{eq:structure_factor_2}) expressing the difference of the magnetic moments of the two sublattices with respect to the $\mathbf{u}_1\mathbf{u}_2\mathbf{u}_3$ reference frame \cite{blume_polarization_1988}.

The extraction of the orbital-to-spin angular momenta ratio $\vert\mathbf{l}\vert/\vert\mathbf{s}\vert$ from the quantity~(\ref{eq:ratio}) appearing in the expression~(\ref{eq:intensities}) for the scattered intensity requires the knowledge of the momentum transfer dependence of the orbital and spin magnetic form factors. Once this is known, an extrapolation to zero momentum transfer can be performed as shown in Fig.~\ref{Orbital_moment}. Although, in general, the form factors depends on the vector $\mathbf{Q}$, an isotropic approximation is usually considered, which only takes into account the dependence on the magnitude $Q$ of the momentum transfer, such that $f_s(\mathbf{Q})\equiv f_s(Q)$ and $f_l(\mathbf{Q})\equiv f_l(Q)$. In this case, the magnetic form factors can be expressed as follows \cite{Prince,Blume1961,langridge_separation_1997}:
\begin{align}
\label{eq:isotropic_form_factors}
f_s(Q)&=\left\langle j_{0}\right\rangle  \nonumber\\
f_l(Q)&=\left\langle j_{0}\right\rangle+\left\langle j_{2}\right\rangle
\end{align}
Here, $\left\langle j_{0}\right\rangle$ and $\left\langle j_{2}\right\rangle$ are radial integrals of the type $\displaystyle
\left\langle j_k\right\rangle_{nl}(Q)=\int_{0}^{\infty}R_{nl}^2(r)j_k(Qr)r^2\,dr$ where $R_{nl}(r)$ is the radial part of the magnetic ion wave function and $j_k(Qr)$ is the spherical Bessel function of order $k$. The radial integrals (espressed as a function of the normalized momentum tranfer $s=Q/4\pi=\sin{\theta}/\lambda$, being $\theta$ the Bragg angle of the magnetic reflection and $\lambda$ the wavelength of the incident x-ray beam) can be approximated by the following \cite{Prince}:
\begin{align}
\label{eq:radial_integral_approx}
\left\langle j_0\right\rangle(s)&=Ae^{-as^2}+Be^{-bs^2}+Ce^{-cs^2}+D \nonumber\\ 
\left\langle j_2\right\rangle(s)&=s^2(A'e^{-as^2}+B'e^{-bs^2}+C'e^{-cs^2}+D')
\end{align}
where the values of the coefficients are tabulated for different oxidation states of each element in Ref.~\onlinecite{Prince}.

\section{Details on the data treatment}\label{sec:data_treatment}
As mentioned in \S~\ref{sec:experimental_NXMS}, as well as the intensity in the $\sigma-\sigma'$ and $\sigma-\pi'$ channels, the total NXMS intensity was also measured in order to correct for the different reflection efficiencies of the PG analyser crystal in the two polarization channels. This mainly originates from the different beam divergence in the vertical and horizontal plane: as a result, the intensity detected in $\sigma-\sigma'$ and $\sigma-\pi'$ will generally be different even in the case of an equal distribution of $\sigma'$ and $\pi'$  polarization. Using the total intensity $I_{tot}$, the measured values of the intensity in $\sigma-\sigma'$ and $\sigma-\pi'$ can be corrected by introducing a compensation factor $f$, which is defined by the following relation:
\begin{align}
\label{eq:compensation_factor}
I_{tot}&=C_{\sigma\sigma'}I_{\sigma\sigma'}^{Measured}+C_{\sigma\pi'}I_{\sigma\pi'}^{Measured}=\nonumber\\
&=C_{\sigma\sigma'}(I_{\sigma\sigma'}^{Measured}+fI_{\sigma\pi'}^{Measured})
\end{align}
Here, $\displaystyle f=\frac{C_{\sigma\pi'}}{C_{\sigma\sigma'}}$ is the ratio of the two arbitrary scale factors which link the intensity measured in the two polarization channels to the total one recorded through the area detector. In the ideal case in which the reflection efficiencies for the two polarization channels are equivalent, only one scale factor would be necessary, which corresponds to having $f=1$.  In practice, the compensation factor $f$ can be extracted using Eq.~(\ref{eq:compensation_factor}) to fit the total intensity as a function of a $360^{\circ}$ rotation of the magnetic field measured with the 2D detector from the $\sigma-\sigma'$ and $\sigma-\pi'$ magnetic field dependences. The measured data can then be corrected by  multiplying the intensity in $\sigma-\pi'$ by $f$: finally, Eq.~(\ref{eq:intensities}) can be used to fit the corrected values.


\bibliography{./PAPER_Carbonates_PRB}

\end{document}